\documentclass[%
 aip,
 sd,%
 amsmath,amssymb,
 reprint,%
]{revtex4-1}

\usepackage{graphicx}
\usepackage{dcolumn}
\usepackage{bm}

\newcommand{\um} {~\mu\mathrm{m}}

\newcommand{\sinc} {\mathrm{sinc}}

\begin{document}

\preprint{AIP/123-QED}

\title[Stress-controlled shear cell]{A stress-controlled shear cell for small-angle light scattering and microscopy}

\author{S. Aime}
\email{stefano.aime@umontpellier.fr}
\author{L. Ramos}
\author{J. M . Fromental}
\author{G. Pr\'{e}vot}
\author{R. Jelinek}
\author{L. Cipelletti}
\email{luca.cipelletti@umontpellier.fr}
\affiliation{Laboratoire Charles Coulomb (L2C), UMR 5221 CNRS-Universit\'{e} de Montpellier, Montpellier, F-France}%

\date{\today}

\begin{abstract}
We develop and test a stress-controlled, parallel plates shear cell that can be coupled to an optical microscope or a small angle light scattering setup, for simultaneous investigation of the rheological response and the microscopic structure of soft materials under an imposed shear stress. In order to minimize friction, the cell is based on an air bearing linear stage, the stress is applied through a contactless magnetic actuator, and the strain is measured through optical sensors. We discuss the contributions of inertia and of the small residual friction to the measured signal and demonstrate the performance of our device in both oscillating and step stress experiments on a variety of viscoelastic materials.
\end{abstract}

\keywords{rheology; light scattering; microscopy; custom shear cell; stress-controlled}
\maketitle

%

\section{Introduction}

Complex fluids such as polymer solutions, surfactant phases, foams, emulsions and colloidal suspensions are ubiquitous in everyday life and industrial applications. Their rheological properties are often of paramount importance both during the production process and for the final user~\cite{larson_structure_1998}. They are also a topic of intense fundamental research in fields as divers as the physics of polymers~\cite{rubinstein_polymer_2003}, the glass transition~\cite{chen_rheology_2010,binder_glassy_2011}, foams dynamics~\cite{cantat_foams_2013}, and biological fluids and tissues~\cite{chen_rheology_2010,janmey_basic_2007}. Conventional rheology is widely used as a powerful characterization tool~\cite{larson_structure_1998}, providing valuable information on the macroscopic mechanical properties of a system. As early as in the 1934 study of shear-induced emulsification by G. I. Taylor~\cite{taylor_formation_1934}, however, it was recognized that coupling rheology to measurements of the sample structure and dynamics at the microscopic scale tremendously increases our insight in the material behavior. Most experiments rely on optical and scattering probes of the microstructure, although other methods have also been introduced, e.g. acoustic velocimetry~\cite{gallot_ultrafast_2013} and nuclear magnetic resonance~\cite{callaghan_rheo-nmr_1999}. Indeed, in the last 40 years a large number of apparatuses have been developed, which use microscopy, static and dynamic light scattering, neutron and X-ray scattering to investigate the microstructure of driven samples. The wide spectrum of topics that have benefitted from such simultaneous measurements demonstrates the importance and success of this approach: a non-exhaustive list includes the investigation of the orientation dynamics and deformation of individual objects such as emulsion drops, polymers, liquid crystals and protein clusters~\cite{taylor_formation_1934,tolstoguzov_deformation_1974,hashimoto_apparatus_1986,van_egmond_time-dependent_1992,di_cola_steady_2008,wieland_studying_2016}, the influence of shear on demixing and critical phenomena~\cite{beysens_light-scattering_1979,lauger_melt_1995}, shear-induced structure distortion and non-equilibrium phase transitions~\cite{ackerson_hard-sphere_1988,yan_shear-induced_1993,linemann_linear_1995,kume_new_1995,vermant_large-scale_1999,varadan_shear-induced_2001,porcar_scaling_2002,ramos_shear_2004,scirocco_effect_2004,cohen_shear-induced_2004,yang_shear-induced_2004,wu_new_2007,kosaka_lamellar--onion_2010,grenard_shear-induced_2011,ramos_structural_2011,lopez-barron_dynamics_2012,akkal_rheo-saxs_2013,gentile_multilamellar_2014,kim_microstructure_2014,lin_biaxial_2014},
the dynamics of foams~\cite{gopal_nonlinear_1995,hohler_periodic_1997} and that of defects in colloidal crystals~\cite{molino_identification_1998,schall_visualization_2004,reinicke_flow-induced_2010,torija_large_2011,tamborini_plasticity_2014}, shear banding~\cite{helgeson_rheology_2009,derks_confocal_2004,angelico_ordering_2010,helgeson_direct_2010,ober_spatially_2011,gurnon_spatiotemporal_2014,chikkadi_shear_2014}  and non-affine deformation in polymer gels and glasses~\cite{basu_nonaffine_2011,2016arXiv160306384N}, and creep and yielding in amorphous, dense emulsions and colloids, and surfactant phases~\cite{hebraud_yielding_1997,petekidis_rearrangements_2002,petekidis_shear-induced_2002,bauer_collective_2006,besseling_three-dimensional_2007,schall_structural_2007,smith_yielding_2007,zausch_equilibrium_2008,chen_rheology_2010, koumakis_yielding_2012,denisov_resolving_2013,lin_far--equilibrium_2013,knowlton_microscopic_2014,sentjabrskaja_creep_2015}.

Previous works may be classified according to the probe method, the deformation geometry and the rheological quantities that are measured. In a first group of experiments, mostly performed in the last 30 years of the past century, small angle static light scattering (SALS) was used to monitor the change of the sample
structure~\cite{tolstoguzov_deformation_1974,beysens_light-scattering_1979,hashimoto_apparatus_1986,ackerson_hard-sphere_1988,van_egmond_time-dependent_1992,lauger_melt_1995,kume_new_1995,vermant_large-scale_1999,varadan_shear-induced_2001,scirocco_effect_2004}. Small angle X-ray scattering (SAXS~\cite{molino_identification_1998,yang_shear-induced_2004,ramos_shear_2004,bauer_collective_2006,di_cola_steady_2008,kosaka_lamellar--onion_2010,torija_large_2011,ramos_structural_2011,akkal_rheo-saxs_2013,wieland_studying_2016}) and neutron scattering (SANS,~\cite{porcar_scaling_2002,helgeson_rheology_2009,angelico_ordering_2010,reinicke_flow-induced_2010,helgeson_direct_2010,lopez-barron_dynamics_2012,kim_microstructure_2014,gurnon_spatiotemporal_2014,gentile_multilamellar_2014} and Ref.~\cite{eberle_flow-sans_2012} and references therein) have also been used~\cite{dhinojwala_micron-gap_1997,denisov_resolving_2013}, although they require large scale facilities and are therefore less accessible than microscopy and light scattering. Quite generally, these experiments were performed under  strain-imposed conditions, often using custom-designed shear cells. Stress measurements (or stress-imposed tests) were available in just a few cases~\cite{hashimoto_apparatus_1986,lauger_melt_1995,vermant_large-scale_1999}, where a commercial rheometer was coupled to a scattering apparatus.

With the development of advanced microscopy methods, in particularly laser scanning confocal microscopy, many groups have developed apparatuses that couple rheology and microscopy~\cite{kume_new_1995,derks_confocal_2004,wu_new_2007,besseling_three-dimensional_2007,schall_structural_2007,zausch_equilibrium_2008,goyon_spatial_2008,besseling_quantitative_2009,chen_microscopic_2010, basu_nonaffine_2011,ober_spatially_2011,grenard_shear-induced_2011,koumakis_yielding_2012,boitte_novel_2013,chan_simple_2013,knowlton_microscopic_2014,tamborini_plasticity_2014,lin_multi-axis_2014,sentjabrskaja_creep_2015}. Apparatuses based on a commercial rheometer usually give access to both the shear stress and the strain~\cite{sentjabrskaja_creep_2015,besseling_quantitative_2009}. With confocal microscopy, both a plate-plate and a cone and plate geometry are possible, since in confocal microscopy the sample is illuminated and the image is collected from the same side. This allows one to avoid any complications in the optical layout due to the wedge-shaped sample volume of the cone and plate geometry. Note that for rotational motion the cone and plate geometry is preferable to the plate-plate one when a uniform stress is required, e.g. for yield stress fluids or in the non-linear regime. Custom shear cells have also been used in conjunction to microscopy. As for scattering-based setups, custom cells are in general strain-controlled, with no measurement of the stress (see however Ref.~\cite{chan_simple_2013} for a notable exception). In spite of this limitation, custom cells may be an interesting option in terms of cost and because they allow valuable features to be implemented: a fine control and a great flexibility on the choice of the surfaces in contact with the sample, the creation of a stagnation plane through counter-moving surfaces~\cite{derks_confocal_2004,wu_new_2007,boitte_novel_2013}, which greatly simplifies the observation under a large applied shear, and the implementation of non-conventional shear geometries, e.g. small channels for investigating confinement effects~\cite{goyon_spatial_2008,besseling_quantitative_2009,ober_spatially_2011} or large-amplitude shear with linearly translating parallel plates~\cite{grenard_shear-induced_2011}.

Real-space microscopy data are unsurpassed in that they provide full knowledge of both the structure and the dynamics of the sample at the particle level. However, microscopy suffers from several limitations: it is quite sensitive to multiple scattering, making turbid samples difficult to study; it requires specifically designed, fluorescently labelled particles if confocal microscopy is to be used; only quite small sample volumes can be imaged; a high resolution, a large field of view and a fast acquisition rate are mutually exclusive, so that a compromise has to be found between these conflicting requirements. Scattering methods, while not accessing the single particle level, do not suffer from these limitations. Furthermore, advanced scattering techniques such as the Time Resolved Correlation~\cite{cipelletti_time-resolved_2003} (TRC) and the Photon Correlation Imaging~\cite{duri_resolving_2009,cipelletti_simultaneous_2013} (PCI) methods can fully capture temporally and spatially varying dynamics, yielding instantaneous coarse-grained maps of the dynamical activity. Thus, scattering techniques are a valuable alternative to real-space methods not only for measuring the sample structure, as in the early works mentioned above, but also to probe its dynamics.

Dynamic light scattering in the highly multiple scattering limit (diffusing wave spectroscopy, DWS~\cite{weitz_diffusing-wave_1993}) has been used since the Nineties of the past century to measure the microscopic dynamics associated with the affine deformation in the shear flow of a simple fluid~\cite{wu_diffusing-wave_1990} and the impact of shear on foam dynamics~\cite{gopal_nonlinear_1995}. In a subsequent series of works, the so-called `echo-DWS' method has been introduced: here, DWS is used to measure the irreversible rearrangements occurring in amorphous viscoelastic solids such as emulsions~\cite{hebraud_yielding_1997}, foams~\cite{hohler_periodic_1997}, or colloidal gels and glasses~\cite{petekidis_rearrangements_2002,petekidis_shear-induced_2002,smith_yielding_2007} subject to an oscillating shear deformation. More recently, space-resolved DWS has been applied to the investigation of the microscopic rearrangements in polymeric solids under compression or elongation~\cite{erpelding_diffusive_2008,2016arXiv160306384N}. Dynamic light scattering in the single scattering regime~\cite{berne_dynamic_1976} (DLS), by contrast, has been much less used as a probe of the microscopic dynamics of a driven system~\cite{tamborini_plasticity_2014,2016arXivarXiv1604.07611}, 
perhaps because single scattering conditions require greater care than multiple scattering ones in designing a scattering apparatus, due to the sensitivity of DLS to the scattering angle at which light is collected and its vulnerability to stray light scattered by any imperfections in the optics~\cite{tamborini_multiangle_2012}. In spite of its greater complexity, DLS has several appealing features, such as its ability to probe the dynamics on a large range of length scales (by varying the scattering angle), the excellent overlap between the probed length scales (from a few tens of nm up to several tens of $\mu$m) and the characteristic structural length scales of most complex fluids, and the possibility to extend it to the X-ray domain (X-photon correlation spectroscopy, XPCS~\cite{madsen_beyond_2010}).

In this paper, we introduce a novel, custom-made shear cell that can be coupled both to a microscope and to a static and dynamic small angle light scattering apparatus, such as that described in Ref.~\cite{tamborini_multiangle_2012}. The shear cell is composed by two parallel plates, one of which can undergo translational motion. In contrast to other custom shear cells previously reported in the literature, for which a shear deformation is imposed and no stress measurement is available~\cite{taylor_formation_1934,tolstoguzov_deformation_1974,van_egmond_time-dependent_1992,ackerson_hard-sphere_1988,gopal_nonlinear_1995,hebraud_yielding_1997,hohler_periodic_1997,molino_identification_1998,varadan_shear-induced_2001,porcar_scaling_2002,petekidis_rearrangements_2002,petekidis_shear-induced_2002,yang_shear-induced_2004,ramos_shear_2004,scirocco_effect_2004,cohen_shear-induced_2004,schall_visualization_2004,wu_new_2007,schall_structural_2007,besseling_three-dimensional_2007,helgeson_rheology_2009,angelico_ordering_2010,chen_microscopic_2010,reinicke_flow-induced_2010,helgeson_direct_2010,torija_large_2011,zausch_equilibrium_2008,basu_nonaffine_2011,koumakis_yielding_2012,boitte_novel_2013,gurnon_spatiotemporal_2014,gentile_multilamellar_2014,knowlton_microscopic_2014,tamborini_plasticity_2014}, our cell is stress-controlled and the strain is accurately measured. Unlike the stress-controlled custom cell of Ref.~\cite{chan_simple_2013} that is optimized for confocal microscopy in an inverted microscope, our cell can be used in both an inverted or upright conventional or confocal microscope, as well as in light scattering experiments in the transmission or backscattering geometry; furthermore, it can be placed either in a vertical or horizontal position. This versatile design opens the way to the full characterization of the rheological behavior of a sample, simultaneously to its structural and dynamical evolution at the microscopic level. In particular, its compact design allows it to be coupled to state-of-the-art static \emph{and} dynamic light scattering setups, a significant improvement over commercially available light scattering adds-on for rheometers, which are limited to static light scattering. 

The rest of the paper is organized as follows: in Sec.~\ref{sec:setup} we briefly present the shear cell, before discussing in Sec.~\ref{sec:setup_characterization} its main features independently of the investigated sample (stress calibration, strain measurement, plate parallelism, effects of inertia and residual friction). Section~\ref{sec:model_fluids} demonstrates the cell performances through a series of tests on model systems, representative of a Newtonian fluid, a perfectly elastic solid, and a viscoelastic Maxwell fluid. In Sec.~\ref{sec:dls}, we illustrate how the cell can be coupled to a dynamic light scattering apparatus, by measuring the microscopic dynamics of a Brownian suspension at rest and under shear. Finally, an overview of the characteristics of the shear cell, with an emphasis on its strengths and limits concludes the paper, in Sec.~\ref{sec:conclusions}.

\section{Setup}
\label{sec:setup}

\begin{figure}[h]
\centering
  \includegraphics[width=\columnwidth,clip]{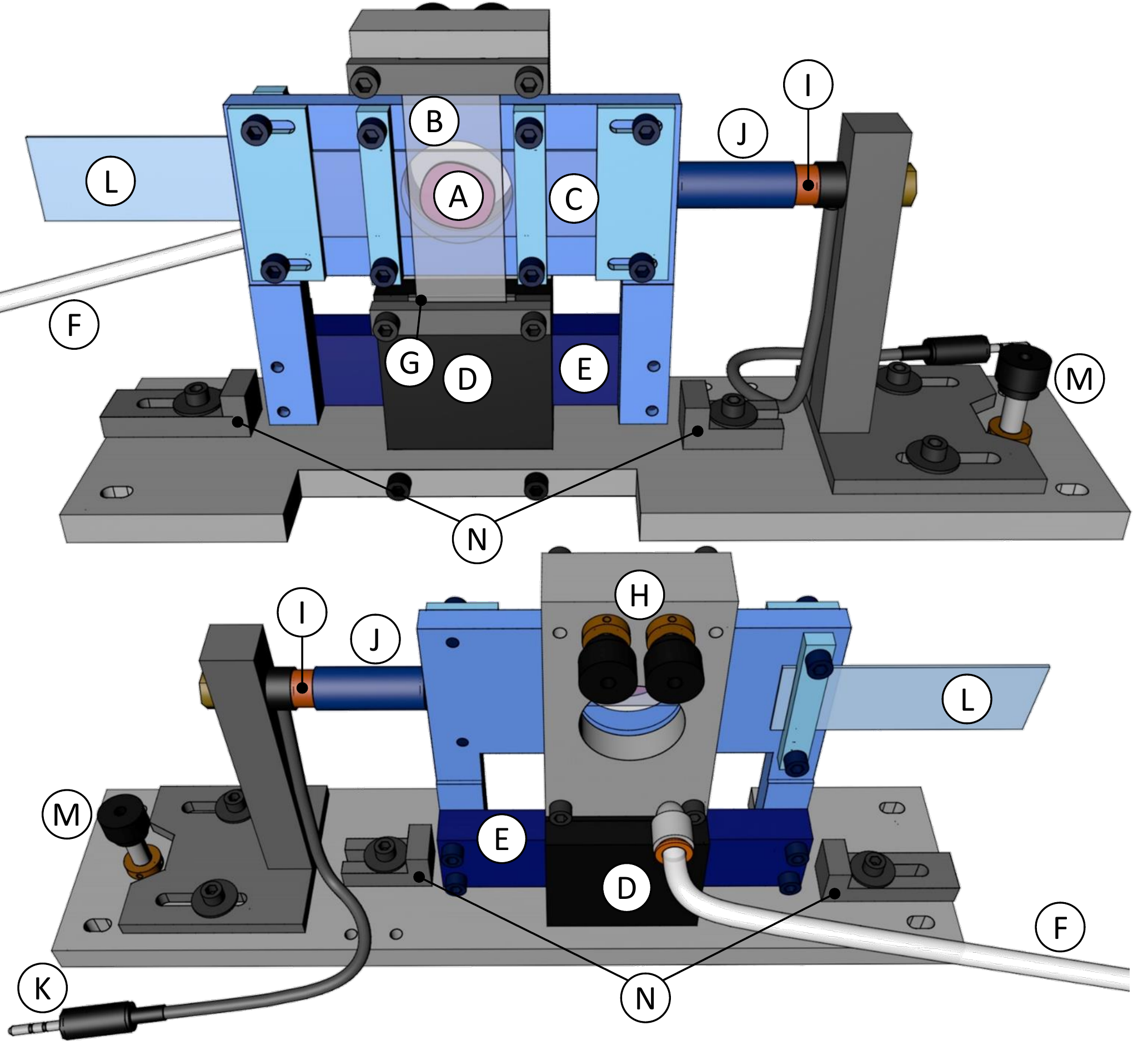}
  \caption{Scheme of the experimental setup. The moving parts are shown in blue. A) Sample; B) fixed glass plate; C) mobile glass plate; D) air bearing - fixed body; E) air bearing - sliding rail; F) air supply; G) gap setting spacer; H) plate alignment screws; I) coil (fixed); J) magnet (mobile); K) cable to PC; L) frosted glass; M) tilt adjustment; N) mechanical stops.}
  \label{fig:shearcell_overview}
\end{figure}

A scheme of the setup is shown in fig.~\ref{fig:shearcell_overview}. The sample is confined between two crossed microscope slides, whose surfaces are separated by a gap $e$, typically of the order of a few hundreds of $\mu$m. The gap can be adjusted using spacers and a set of precision screws to ensure the parallelism of the plates, as described in more detail in Sec.~\ref{sec:plates}. One of the two slides is fixed to an optical table, while the second one is mounted on a mobile frame that slides on a linear air bearing. The air bearing avoids any mechanical contact between the sliding frame and the optical table, therefore minimizing friction. In order to avoid spurious contributions due to gravity, the horizontal alignment of the air bearing rail can be adjusted to within a few $10^{-5}$ rad using a finely threaded differential screw. To impose the desired shear stress, a controlled force is applied to the mobile frame using a magnetic actuator (see fig.~\ref{fig:shearcell_overview}), powered by a custom-designed, computer-controlled current generator. A contactless optical sensor measures the displacement of the moving frame and thus the sample strain. The implementation of the strain sensor is discussed in Sec.~\ref{sec:strain}.

The linear air bearing (model RAB1, from Nelson Air Corp.) is guided by air films of thickness $\sim 10~\mu$m, which confine its motion in the longitudinal $x$ direction. The viscous drag of the air films can typically be neglected, whereas a small residual friction, most likely due to dust particles, is seen when a force smaller than $1$ mN is applied, as it will be shown in Sec.~\ref{sec:empty}.
The maximum linear travel of the system, set by the length of the moving plate, is 75 mm. However, we typically restrain the travel distance to 5-10 mm, over which the imposed force is virtually independent of position (see Sec.~\ref{sec:stress}). This typically corresponds to a strain $\gamma \le 10$. The nominal straightness of the sliding bar (provided by the manufacturer) is better than $1~\mu$m over the whole travel length, whereas the nominal thickness of the plates is constant to within $10~\mu$m throughout their entire surface. Therefore, the gap $e$ can be considered to be uniform over the typical working distance. More details on the plates parallelism will be given in Sec.~\ref{sec:plates}.
The sliding part of the setup has a relatively small mass, $M \simeq 335$ g, including the mass of the sliding bar itself, $288.0$ g. This allows inertial effects to be kept small, as shown in Sec.~\ref{sec:model_fluids}.

\section{Setup characterization}
\label{sec:setup_characterization}

In this Section, we describe the characterization of the empty shear cell, discussing in particular the force-current calibration of the stress actuator, the optical measurement of the strain, the control and measurement of the plate parallelism, and the effects of inertia and residual friction.

\subsection{Calibration of the applied force}
\label{sec:stress}

\begin{figure}[h]
\centering
  \includegraphics[width=\columnwidth,clip]{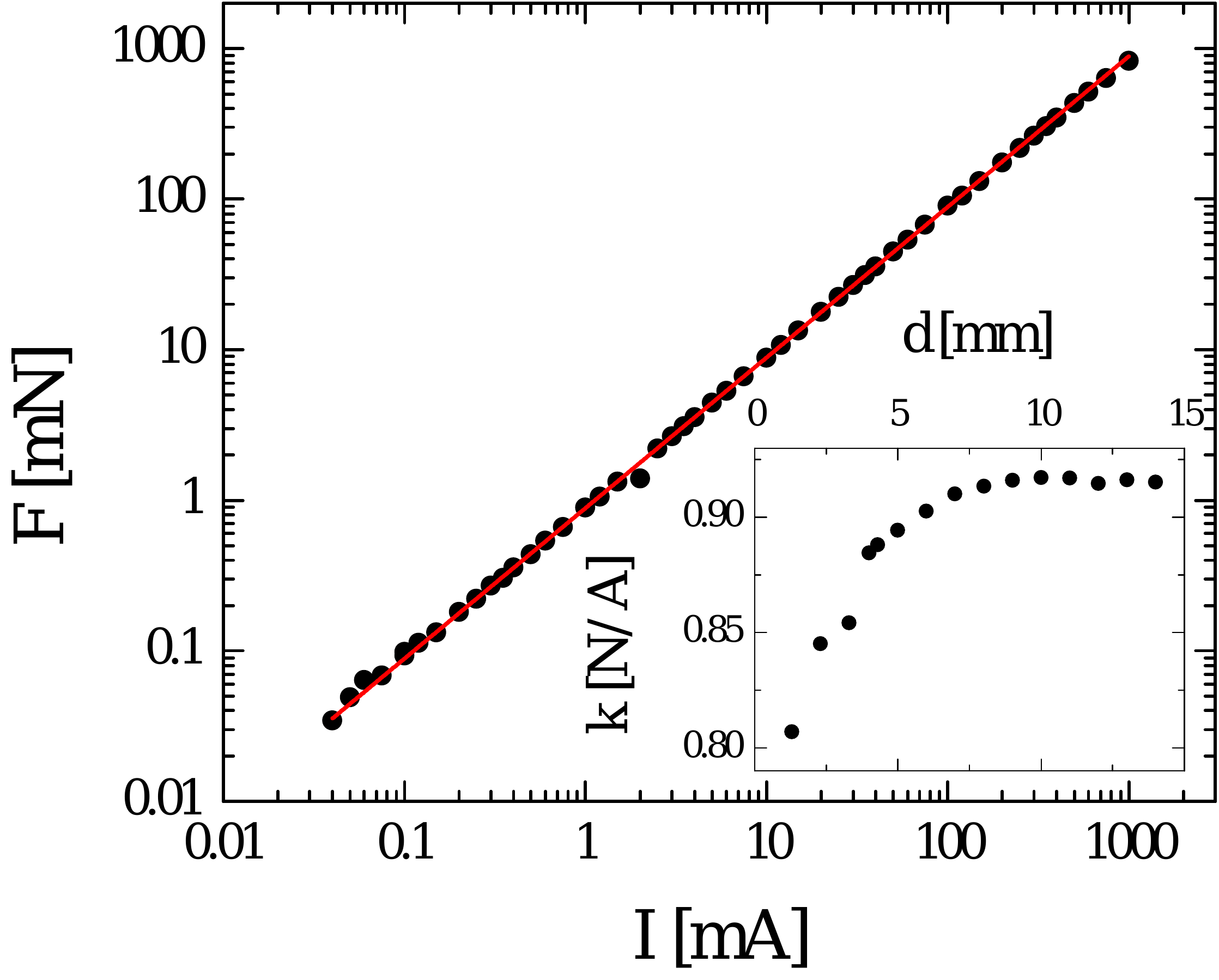}
  \caption{Calibration of the magnetic actuator: force (measured by a precision balance) as a function of the applied current. The symbols are the experimental data and the line is the best fit of a straight line through the origin, yielding a proportionality constant $k=(0.916 \pm 0.003)~\mathrm{NA}^{-1}$. The upper limit of the imposed current is set by the coil specifications, whereas the smallest current chosen here is limited by the balance precision. Inset: variation of $k$ with the relative axial position between the magnet and the coil, $d$. The main plot has been measured for $d=7$ mm.}
 \label{fig:magnetic_actuator}
\end{figure}

The shear stress imposed to the sample is controlled using a magnetic actuator (Linear Voice Coil DC Motor LVCM-013-032-02, by Moticont), consisting of a permanent magnet and a coil winded on an empty cylinder. The magnet is fixed to the sliding part of the apparatus; it moves with no contact within the empty cylinder, which is mounted on the fixed part of the apparatus. The device can be easily controlled via a PC, which allows one to synchronously apply a given stress and acquire strain and optical data. Due to the small coil resistance ($5.9$ $\Omega$) and inductance ($1$ mH), the magnetic actuator response is very fast and its characteristic time (less than $1$ ms) will be neglected in the following.
The force exerted by the voice coil is proportional to the current fed to the device. We use a custom-designed current generator, which outputs a controlled current $I$ that can be adjusted over a very large range by selecting the appropriate full scale value, between  $10^{-5}~\mathrm{A}$ and $1~\mathrm{A}$, in steps of one decade. The current generator noise is smaller than $10^{-4}$ of the full scale. The current generator is in turn controlled through a voltage input: by feeding different voltage waveforms to the current generator, it is therefore possible to impose a stress with an arbitrary time dependence. In our implementation, the voltage signal is generated by a D/A card (USB-6002 by National Instruments) controlled by a PC.
We calibrate the voice coil by measuring the force, $F$, resulting from the imposed current. The force is measured using a balance, with a precision of $1$ mg. The $F$ vs. $I$ curve is shown in fig.~\ref{fig:magnetic_actuator}. $F$ is remarkably proportional to $I$ over 4.5 decades, with a proportionality constant $k=(0.916 \pm 0.003)~\mathrm{NA}^{-1}$.

An important issue in designing the shear cell apparatus is the requirement that the applied stress be constant regardless of the resulting strain. To check this point, we have measured several calibration curves similar to that shown in fig.~\ref{fig:magnetic_actuator}, each time slightly changing $d$, the relative axial position between the magnet and the coil, with $d=0$ the position of the magnet when it is fully inserted in the coil. The $d$ dependence of the proportionality constant $k$ is shown in the inset of fig.~\ref{fig:magnetic_actuator}. In the range $7~\mathrm{mm} < d < 15~\mathrm{mm}$, we find that $k$ changes by less than 0.5\%. In our experiments we typically work in this range, thus ensuring that the applied force is essentially independent of sample deformation. In order to convert the applied force to a stress value, we measure the surface $S$ of the sample by taking a picture of the shear cell after loading it. The surface is calculated from the image using standard image processing tools. Typical values of $S$ and its uncertainty are 250 mm$^2$ and a few mm$^2$, respectively. When taking into account both the uncertainty of the $F-I$ calibration ($0.3\%$) and that on $S$, the stress $\sigma = F/S$ is known to within about 1\%.

\subsection{Strain measurement}
\label{sec:strain}

\begin{figure}[h]
\centering
  \includegraphics[width=\columnwidth,clip]{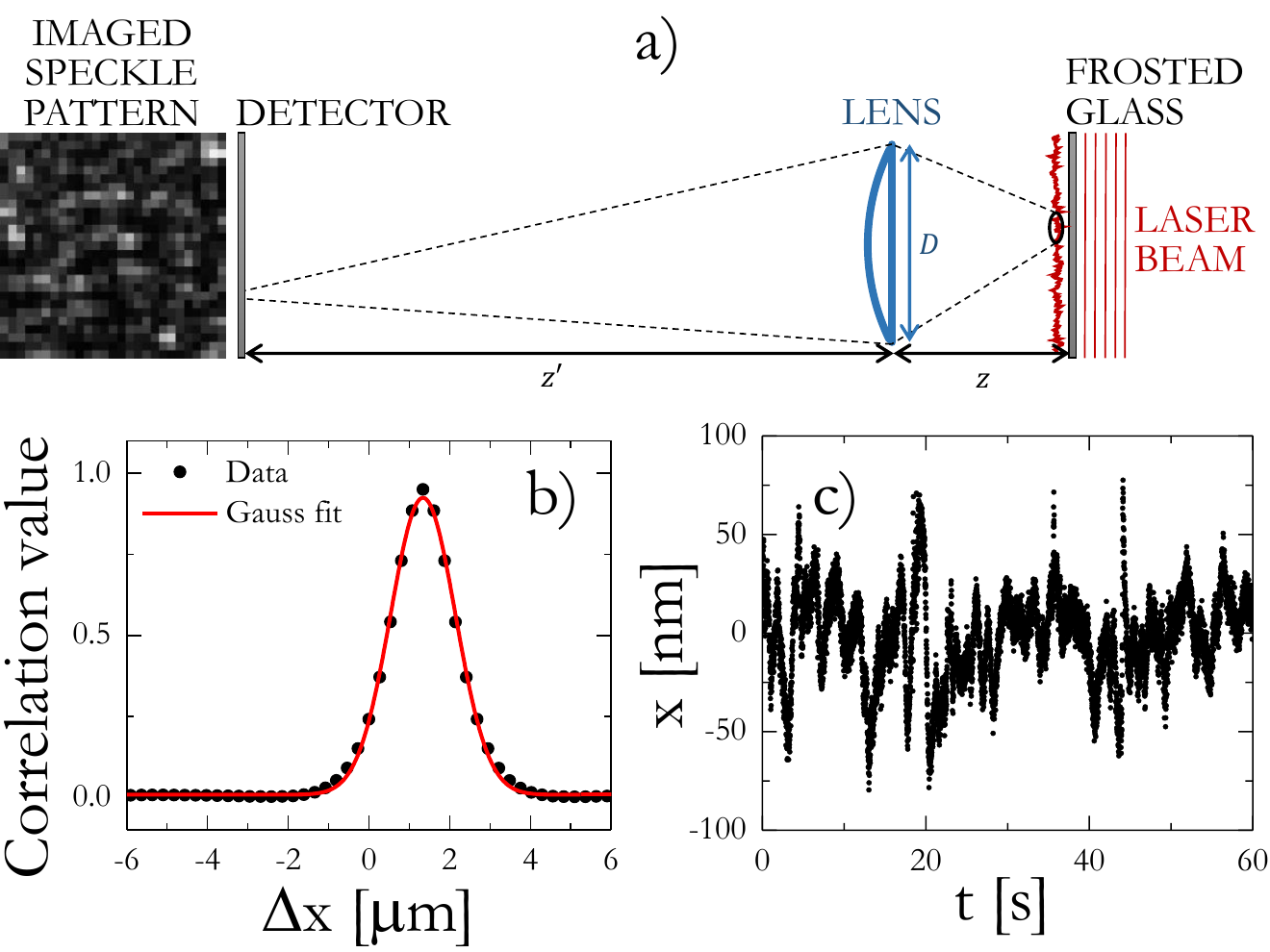}
  \caption{(a) Optical scheme of the speckle imaging system for measuring the strain: $D=12.7$ mm, $z=20$ mm and $z^\prime=410$ mm. (b) Cut along the $x$ direction of the crosscorrelation between two speckle images taken at time $t$ and $t'$. The peak position, $\Delta x = 1.34~\mu \mathrm{m}$, corresponds to the translation of the mobile part of the cell between $t$ and $t'$. The line is a Gaussian fit to the peak, yielding a speckle size (in the sensor plane) $\sigma_{sp}=16.5~\mu$m. (c) Time dependence of the measured displacement while the cell is at rest. The standard deviation $\sigma_n = 23$ nm of the measured position yields an estimate of the measurement error.}
  \label{fig:SpeckleMeter_schema}
\end{figure}

In order to measure the displacement of the mobile frame and hence the sample strain, we use two different optical methods, based on a commercial sensor and on a scattering technique, respectively. The former is a commercial laser position sensor (model IL-S025, by Keyence) that can acquire and send to a PC up to $1000$ points per second with a nominal precision of $1~\mu$m. This sensor is very convenient to follow large, fast deformations. However, its precision sets a lower bound on the error on the strain measurements of at least $0.5 \%$ for a typical gap $e = 200~\mu\mathrm{m}$; moreover, we find that the sensor output tends to artifactually drift over time. When more precise, more stable strain measurements are required, we use a custom optical setup, inspired by~\cite{cipelletti_simultaneous_2013} and schematized in fig.~\ref{fig:SpeckleMeter_schema}a. A frosted glass is fixed to the mobile frame of the shear cell. A laser beam illuminates the frosted glass, which is imaged on a CMOS camera (BU406M by Toshiba Teli Corp). The camera has a $2048 \times 2048$ $\mathrm{pixel}^2$ detector, with a pixel size of $5.5~\mu$m; its maximum acquisition rate at full frame is $90$ Hz. The image is formed by a plano-convex lens with a focal length $f = 19$ mm, placed in order to achieve a high magnification, $m=20.5$, such that the pixel size corresponds to $268$ nm on the frosted glass. Due to the coherence of the laser light, the image consists of a highly contrasted speckle pattern (see fig.~\ref{fig:SpeckleMeter_schema}a), which translates as the mobile part of the cell, and hence the frosted glass, is displaced. Using image cross-correlation techniques~\cite{tokumaru_image_1995} the speckle drift and thus the sample shear can be measured. In practice, a speckle image acquired at time $t$ is spatially cross-correlated with an image taken at a later time $t'$. The position of the peak of cross-correlation as a function of the spatial shift yields the desired displacement between times $t$ and $t'$. An example is shown in fig.~\ref{fig:SpeckleMeter_schema}b, which shows a cut of the cross-correlation function along the $x$ direction. The line is a Gaussian fit to the peak: when taking into account the magnification, the fit yields a speckle size $\sigma_{sp}=16.5~\mu$m in the plane of the sensor. The speckle size is controlled by the diameter $D$ of the lens and the system magnification; it has been optimized in order to correspond to a few pixels, which minimizes the
noise on the cross-correlation~\cite{viasnoff_multispeckle_2002}. 

To estimate the typical noise on the measurement of the displacement, we take a series of full-frame images of the speckle pattern over a period of $60$ s, while keeping the frosted glass immobile. The displacement with respect to the first image ($t=0$), as measured from the position of the cross-correlation peak, is shown in fig.~\ref{fig:SpeckleMeter_schema}c. The displacement fluctuates around a zero average value; the standard deviation of the signal over the full acquisition time, $\sigma_n$, may be taken as an estimate of the typical noise on the measured position. We find $\sigma_n = 23$ nm, more than 40 times smaller than the nominal precision of the commercial laser sensor. For a typical gap $e = 200~\mu$m it is therefore possible to reliably measure strains as small as $10^{-4}$. The main limitations of this technique are its time resolution and the largest displacement speed that can be directly measured. The former is limited by the camera acquisition rate and the image processing time: in practice, the position can be sampled at a maximum rate of about 50 Hz. The latter is limited by the acquisition rate and size of the field of view, $l_v \approx 0.5~\mathrm{mm}$: if the frosted glass translates by more than $l_v$ between two consecutive images, the speckle pattern is completely changed and no cross-correlation peak is observed. In practice, displacement speeds as high as $1 \mathrm{mm~s}^{-1}$ can be measured, corresponding to strain rates up to $\dot{\gamma} = 2~s^{-1}$ or $\dot{\gamma} = 5~s^{-1}$ for gaps of $500 \um$ and $200 \um$, respectively. Since the commercial sensor and our custom device have complementary strengths and limitations, we typically use both of them simultaneously. A final point concerns the conversion of the displacement to strain units, for which the gap thickness is required. We determine it after adjusting the plates parallelism (see Sec.~\ref{sec:plates} below), by placing the empty shear cell under an optical microscope and by measuring the vertical displacement required to focus the inner surface of the fixed and moving plates, respectively.

\subsection{Adjusting the parallelism between the moving and fixed plates}
\label{sec:plates}

\begin{figure}[h]
\centering
  \includegraphics[width=\columnwidth,clip]{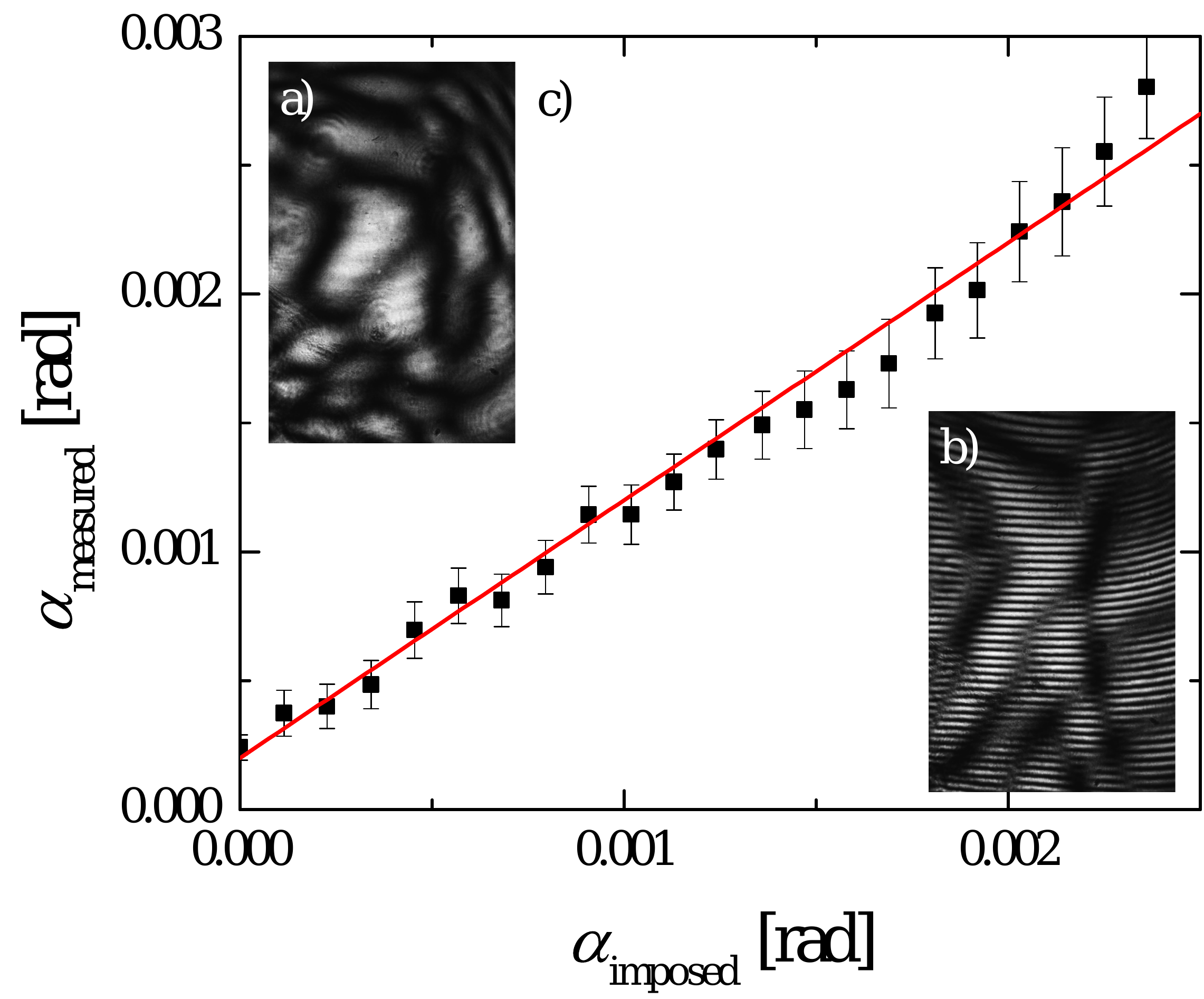}
  \caption{(a) and (b): images of the interference fringes observed when the plates are parallel or tilted by an angle $\alpha_{imposed} = 8\times 10^{-4}$ rad, respectively. The size of the field of view is $15 \times 22~\mathrm{mm}^2$. (c) Measured tilt angle as a function of the manually imposed tilt angle. The symbols are the data and the line is a linear fit with slope $1$.}
  \label{fig:plates_parallelism}
\end{figure}

The parallelism between the two plates is tuned using two differential screws (model DAS110, by ThorLabs), which control the position of the upper side of the fixed plate (see the pictures in fig.~\ref{fig:shearcell_overview}). The quality of the alignment is checked by visualizing the interference fringes formed by an auxiliary laser beam reflected by the two inner surfaces of the plates. A change in the local separation between the two plates modifies the relative phase of the two reflected beams, thus changing the fringe pattern. When the plates are parallel, no fringes should be visible. The fringes are imaged on a CMOS camera (DMM 22BUC03-ML, by The Imaging Source, GmbH) with a $744 \times 480~\mathrm{pixel}^2$ detector, the pixel size being $6~\mu$m, corresponding to $31~\mu$m in the sample plane (magnification factor $m=0.19$). Figure~\ref{fig:plates_parallelism}b shows the fringe pattern observed when a tilt angle of $8\times 10^{-4}~\mathrm{rad}$ between the two plates is imposed on purpose, using the two differential screws. In addition to the finely spaced fringes, due to the tilt, some larger-scale, irregular fringes are also observed, due to slight deviations of the plates from perfectly flat surfaces. When optimizing the alignment (fig.~\ref{fig:plates_parallelism}a), only the irregular, large-scale fringes are visible. To gauge the precision with which the plates can be aligned, we impose a series of increasingly large tilt angles $\alpha_{imposed}$, acting on the two differential screws. $\alpha_{imposed}$ is estimated from the length of the plate, the nominal thread of the screws, and the imposed screw rotation. For each $\alpha_{imposed}$, we measure the tilt angle $\alpha_{measured} = N\lambda/2L$, where $N$ is the number of fine fringes observed over a distance $L$ and $\lambda = 633~\mathrm{nm}$ is the laser wavelength. Figure~\ref{fig:plates_parallelism}c shows the measured tilt angle as a function of the imposed one: the data are very well fitted by a straight line with unit slope and a small offset, $\sigma_\alpha \sim 0.3~\mathrm{mrad}$, most likely due to the difficulty of finding the optimum alignment since the plate surfaces are not perfectly flat. We conclude that the plates can be tuned to be parallel to within a tilt angle of about 0.3 mrad, a value comparable to or even better than for commercial rheometers~\cite{rodriguez-lopez_using_2013}. Once the plates have been aligned, they can be removed and placed again in the setup (e.g. for cleaning them) with no need to realign the setup. In this case, we checked that the parallelism is preserved to within 1 mrad.

\subsection{Effects of inertia and friction}
\label{sec:empty}

\begin{figure}[h]
\centering
  \includegraphics[width=\columnwidth,clip]{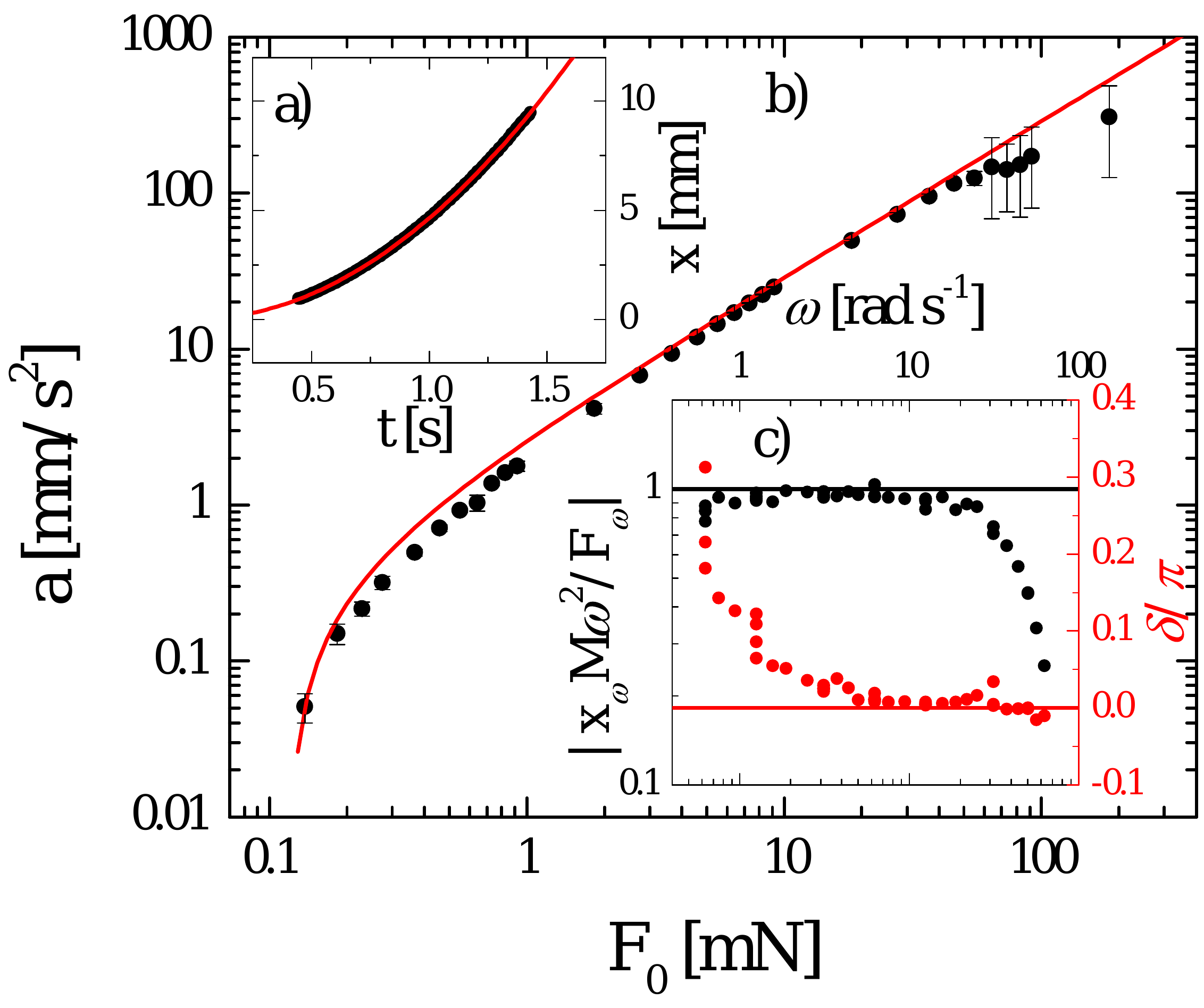}
  \caption{(a) time-dependent translation $x(t)$ following the application of a step force $F_0=0.9$ mN to the empty cell. The line is a quadratic fit. (b)  Acceleration of the empty cell \textit{vs.} the amplitude of the step force applied. Symbols are the data point and the line is the best fit of an affine law, $a=\frac{F_0-F_{fr}}{M}$, with $M=335$ g the mass of the sliding frame, yielding a residual friction force $F_{fr}=0.12$ mN as the only fitting parameter. (c) Normalized amplitude (black, left axis) and phase (red, right axis) of the oscillating part of the response to a cosinusoidal external force. The symbols are the data points and the lines are the theoretical behavior.}
  \label{fig:empty}
\end{figure}

Inertia and friction effects may lead to spurious results if they are neglected in the modelling of the setup response. Both effects are best characterized by measuring the cell motion under an applied force in the absence of any sample. Under these conditions, the equation of motion for the moving part of the cell reads:
\begin{equation}
    M\ddot{x}(t)=F^{(ext)}(t)+F_{fr}\left[x(t), \dot{x}(t)\right] \,,
    \label{eqn:force_balance_empty}
\end{equation}
where $M$ is the mass of the moving part, $x$ its position (with $x(0)=0$ the position before applying any force), $F^{(ext)}$ the (time-dependent) force applied by the electromagnetic actuator, and $F_{fr}$ a friction term that may depend on both the position (in the case of solid friction) and the velocity (for viscous-like friction). To characterize the friction term, we measure $x(t)$ after imposing a step force, $F^{(ext)} = F_0\Theta(t)$, with $\Theta(t)$ the Heaviside function. Figure~\ref{fig:empty}a shows that $x(t)$ follows remarkably well a quadratic law, suggesting that the friction contribution, if any, is independent of both $x$ and $\dot{x}$. We perform the same experiment for a variety of applied forces and extract, for each $F_0$, the acceleration $a$ from a quadratic fit to $x(t)$. The results are shown in fig.~\ref{fig:empty}b: while at large $F_0$ one finds $a\propto F_0$, as expected if friction is negligible, for $F_0 < 1~\mathrm{mN}$ the data clearly depart from a proportionality relationship, suggesting that friction reduces the effective force acting on the moving part of the cell. The line in fig.~\ref{fig:empty}b is a fit to the data of the affine law $a = (F_0-F_{fr})/M$, with $M$ fixed to $335~\mathrm{g}$ , yielding $F_{fr} = 0.12~ \mathrm{mN}$. The fit captures very well the behavior of $a$ over 4 decades, showing that a static friction term accounts satisfactorily for the experimental data. Slight deviations are observed for large forces, presumably due to the large uncertainty in the measurement of the displacement, a consequence of the very fast motion in this regime. Some small deviations are also observed at low forces, possibly because the friction term may slightly deviate from the simple position- and velocity-independent expression that we have assumed. Since the sample surface is typically $A \sim 2~\mathrm{cm}^2$, the friction term sets a lower bound of the order of 1 Pa on the stress that may be applied in our apparatus.

To probe the frequency dependence of the setup response, we apply an oscillatory force to the empty cell, $F(t)=\mathrm{Re} \left(\tilde{F}_{\omega} e^{i \omega t} \right )$, where the tilded variables are complex quantities. 
Assuming $x(0) = \dot{x}(0) = 0$ and neglecting friction, the expected response is
\begin{equation}
	\tilde{x}(t)= \tilde{x}_\omega \left( e^{i \omega t} - 1 - i \omega t \right)\,,
	\label{eqn:inertial_general_sol}
\end{equation}
with $\tilde{x}_{\omega}=-\tilde{F}_{\omega}/(M \omega^2)$. Note that, because the physical equation of motion is given by the real part of Eq.~\ref{eqn:inertial_general_sol}, inertial effects yield either a nonzero offset $\tilde{F}_{\omega}/(M\omega^2)$ (for a cosine-like applied force), or a nonzero drift velocity $i\tilde{F}_{\omega}/(M\omega)$ (for a sinusoidal force), or both of them if the real and imaginary parts of $\tilde{F}_{\omega}$ are nonzero. Thus, in oscillatory shear experiments inertia effects add a constant or linearly growing strain to the sample, whose contribution may or may not be negligible depending on the sample rheological properties and the excitation frequency. To test Eq.~\ref{eqn:inertial_general_sol}, we apply a cosinusoidal force to the empty cell, varying the driving frequency from $0.75~\mathrm{rad~s}^{-1}$ to $62.8~\mathrm{rad~s}^{-1}$. For each $\omega$, we adjust $F_{\omega}$ so as to keep the amplitude of the oscillations roughly constant and measure both the modulus and the phase of $\tilde{x}_{\omega}$ from the time dependence of the displacement in the stationary regime. Figure~\ref{fig:empty}c shows that the magnitude of the oscillations has the expected value up to $4~\mathrm{rad~s}^{-1}$, beyond which it drops, due to the frequency dependence of the response of the current generator and magnetic actuator. The phase of $\tilde{x}_{\omega}$ is close to zero, as expected, except for the lowest frequencies, for which the applied force is too small for friction effects to be neglected. 

\section{Tests with model samples}
\label{sec:model_fluids}

Having characterized the response of the empty cell, we test its performances on standard rheological samples in oscillatory and step stress experiments. We use a purely viscous fluid, a purely elastic solid, and a viscoelastic material well described by a Maxwell model. For each sample, we briefly discuss the expected behavior in the presence of inertia and compare it to the outcome of experiments, in order to assess the sensitivity and limits of our setup. For the sake of simplicity, we will not introduce the friction term in the equations, but its effects will be highlighted in discussing the experiments.

When a sample is loaded in the shear cell, Eq.~\ref{eqn:force_balance_empty} has to be modified in order to take into account the force exerted by the sample on the cell, $F^{(s)}(x, \dot{x})$:
\begin{equation}
    M\ddot{x}(t)=F^{(ext)}(t)+F_{fr}\left[x(t), \dot{x}(t)\right] +F^{(s)}\left[x(t), \dot{x}(t)\right]\,.
    \label{eqn:force_balance}
\end{equation}
In principle, $F^{(s)}(x, \dot{x})$ can be divided into an elastic ($x$-dependent) and a viscous ($\dot{x}$-dependent) parts.
To recast Eq.~\ref{eqn:force_balance} in a form suitable to describe the sample rheological properties, we divide both sides by the sample surface $A$ and express position and velocity as strain $\gamma = x/e$ and strain rate $\dot{\gamma}$, respectively:
\begin{equation}
    I\ddot{\gamma}(t)+\sigma\left[\gamma(t), \dot{\gamma}(t)\right]=\sigma^{(ext)}(t)
    \label{eqn:force_balance_unitarea}
\end{equation}
Here, $I=eM/A$ is the inertia term, $\sigma^{(ext)}$ the applied stress and $\sigma = -F^{(s)}/A$ the sample stress, which has opposite sign with respect to $F^{(s)}$ since we adopt the usual notation where $\sigma$ is the stress exerted by the setup on the sample. This is the general equation that has to be solved given a constitutive equation for $\sigma\left[\gamma(t), \dot{\gamma}(t)\right]$, an experimental protocol (i.e. the temporal profile of $\sigma^{(ext)}$), and the initial conditions. In the following, we will systematically assume $\gamma(0)=\dot{\gamma}(0)=0$, and will only consider either creep tests, $\sigma^{(ext)}(t) =\sigma_0\Theta (t)$, or oscillatory stresses, $\sigma^{(ext)}(t)=\tilde{\sigma}_\omega e^{i \omega t}$.

\subsection{Newtonian fluid}

\subsubsection{Theoretical background.}

For a Newtonian fluid of viscosity $\eta$, $\sigma=\eta \dot{\gamma}$. Equation~\ref{eqn:force_balance_unitarea} therefore reads:
\begin{equation}
	I\ddot{\gamma} + \eta \dot{\gamma} = \sigma^{(ext)}\,,
\label{eqn:viscous}
\end{equation}
which, for a step stress test with stress amplitude $\sigma_0$ , has as solution
\begin{equation}
    \gamma(t) = \dot{\gamma}_\infty \left[t + \tau_v \left( e^{-\frac{t}{\tau_v}} - 1 \right)\right] \,,
    \label{eqn:viscous_motion}
\end{equation}
where $\tau_v=I/\eta$ is a characteristic time arising from the interplay between inertial and viscous effects and $\dot{\gamma}_\infty = \sigma_0/\eta$. For $t \ll \tau_v$, inertia dominates and $\gamma \approx \sigma_0 t^2/(2I)$. In the opposite limit $t \gg \tau_v$, the usual simple viscous flow $\gamma = \dot{\gamma}_\infty t$ is recovered. Typical values of the crossover time may be estimated from $\tau_v~\mathrm{[s]} \approx (2 \eta \mathrm{[Pa~s]})^{-1}$, where we have assumed $M=0.3$ kg, $e=300~\mu$m and $A=2~\mathrm{cm}^2$. Thus, the inertial regime lasts less than 1 s if the viscosity exceeds $0.5~\mathrm{Pa~s}^{-1}$.

For an oscillatory stress, the solution to Eq.~\ref{eqn:viscous} is
\begin{equation}
	\gamma(t) = \tilde{\gamma}_\omega \left[e^{i \omega t}-1 + i \omega \tau_v \left(e^{-\frac{t}{\tau_v}}-1\right)\right]\,,
\label{eqn:viscousosc}
\end{equation}
with
\begin{equation}
   \tilde{\gamma}_\omega = -\frac{\tilde{\sigma}_\omega}{\eta \omega} \frac{i + \omega \tau_v}{1 + \omega^2\tau_v^2}\,.
	\label{eqn:viscous_oscill_amplitude}
\end{equation}

The frequency-dependent complex modulus extracted from the oscillating part of $\gamma(t)$, $G^* = \tilde{\sigma}_\omega/\tilde{\gamma}_\omega$, has magnitude and phase given by
\begin{equation}
\label{eqn:viscous_moduli}
\begin{array}{r@{}l}
		|G^*| = \eta \omega \sqrt{1+\omega^2\tau_v^2} \\
		\arg G^* =\delta =  \arctan \frac{1}{\omega\tau_v}\,.
\end{array}
\end{equation}

At short times  $t \ll \tau_v$, viscous dissipation is always negligible and $\gamma(t) \approx \tilde{\gamma}_\omega\left( e^{i\omega t}-1-i \omega t\right)$, as in Eq.~\ref{eqn:inertial_general_sol}, which was derived for an empty cell. At larger times, Eq. ~\ref{eqn:viscousosc} simplifies to $\gamma(t) \approx i \tilde{\sigma}_\omega/(\eta \omega)\left[1 - e^{i\omega t}/\left(1+i\omega t\right)\right]$: in the high frequency regime $\omega\tau_v \gg 1$ inertial effects are still relevant, but in the opposite regime $\omega\tau_v \ll 1$ the usual behavior for a viscous fluid is recovered: the cell oscillates around a negligible equilibrium position $\tilde{\gamma}_\omega \omega \tau_v$, with an amplitude $\tilde{\gamma}_\omega\approx -i \tilde{\sigma}_\omega/(\eta \omega)$, and the strain lags the stress by an angle $\delta = \pi/2$.

\subsubsection{Experimental tests}

\begin{figure}[h]
\centering
  \includegraphics[width=\columnwidth,clip]{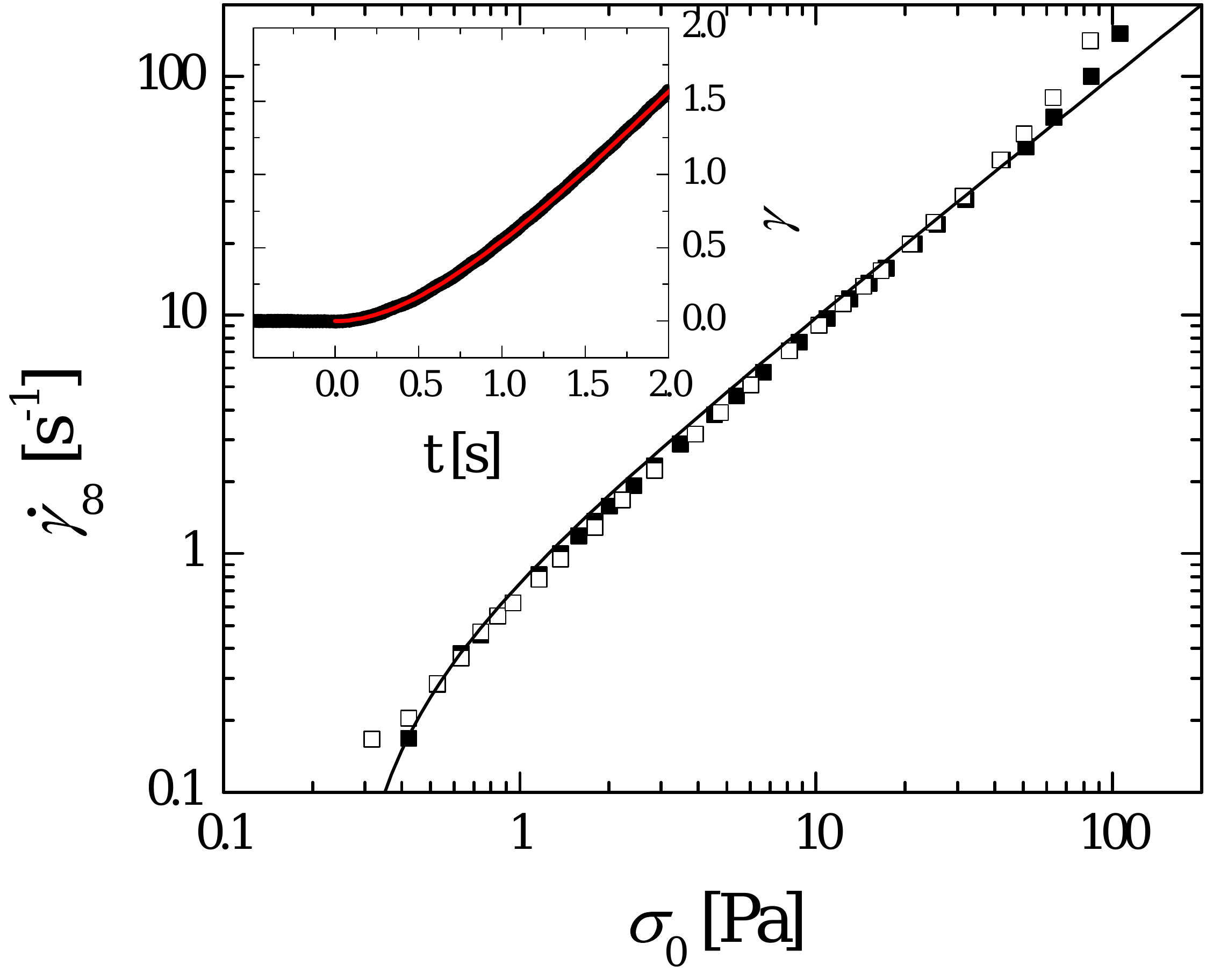}
  \caption{Main plot: steady shear rate $\dot{\gamma}_\infty$ as a function of the applied stress in a step stress test, for a silicon oil with nominal viscosity $\eta=1.02~\mathrm{Pa~s}$. Empty and filled symbols refer to two independent measurements. The line is a fit of  $\dot{\gamma}_\infty = (\sigma - \sigma_{fr})/\eta$ where the offset $\sigma_{fr}\sim 0.25$ Pa is due to friction. Inset: time evolution of the strain following a step stress of amplitude $\sigma_0 = 1.58$ Pa.  The symbols are the data and the line is a fit of Eq.~\ref{eqn:viscous_motion}, with $\tau_v = 0.7~\mathrm{s}$ and $\dot{\gamma}_\infty = 1.18~\mathrm{s}^{-1}$, in good agreement with $(\sigma_0-\sigma_{fr})/\eta = 1.27~\mathrm{s}^{-1}$ as obtained from the nominal viscosity.}
  \label{fig:viscous_step}
\end{figure}

Measurements for step-stresses have been performed with 3 different silicon oils, with viscosity $1.02$, $11.98$ and $91.68$ Pa~s respectively (nominal values at $25^\circ$C). As an example, we show in fig.~\ref{fig:viscous_step} results for the less viscous oil, for which the applied stress is varied between $0.3$ and $100$ Pa. The inset shows the time evolution of the strain following the application of a step stress with $\sigma_0= 4.5$ Pa. At large times, a pure viscous flow is found, $\gamma(t) = \dot{\gamma}_\infty t$. Deviations at short times from this linear behavior are due to inertia. The data are in excellent agreement with the theoretical expression, Eq.~\ref{eqn:viscous_motion} (line), where  $\tau_v=eM/(\eta A)$ and $\dot{\gamma}_\infty$ are fitting parameters. We plot in the main figure the steady-state shear rate $\dot{\gamma}_\infty$ vs $\sigma_0$, obtained by fitting $\gamma(t)$ for all applied stresses. At large stresses, data are in excellent agreement with $\dot{\gamma} = \sigma_0/\eta$ using the nominal value of the viscosity, whereas friction causes deviations from this linear dependence for the smallest stresses. Note that friction becomes relevant for $\sigma_0 \lesssim 1~\mathrm{Pa}$, in agreement with the lower bound on the stress estimated in Sec.~\ref{sec:empty}. Data over the whole range of applied stresses are very well accounted for by Eq.~\ref{eqn:viscous_motion} (line in fig.~\ref{fig:viscous_step}), with $\tau_v$ as the only fitting parameter and using the nominal viscosity. The fit yields $\tau_v=0.7~\mathrm{s}$, in good agrement with the expected value $eM/(\eta A) = 0.8~\mathrm{s}$.

\begin{figure}[h]
\centering
  \includegraphics[width=\columnwidth,clip]{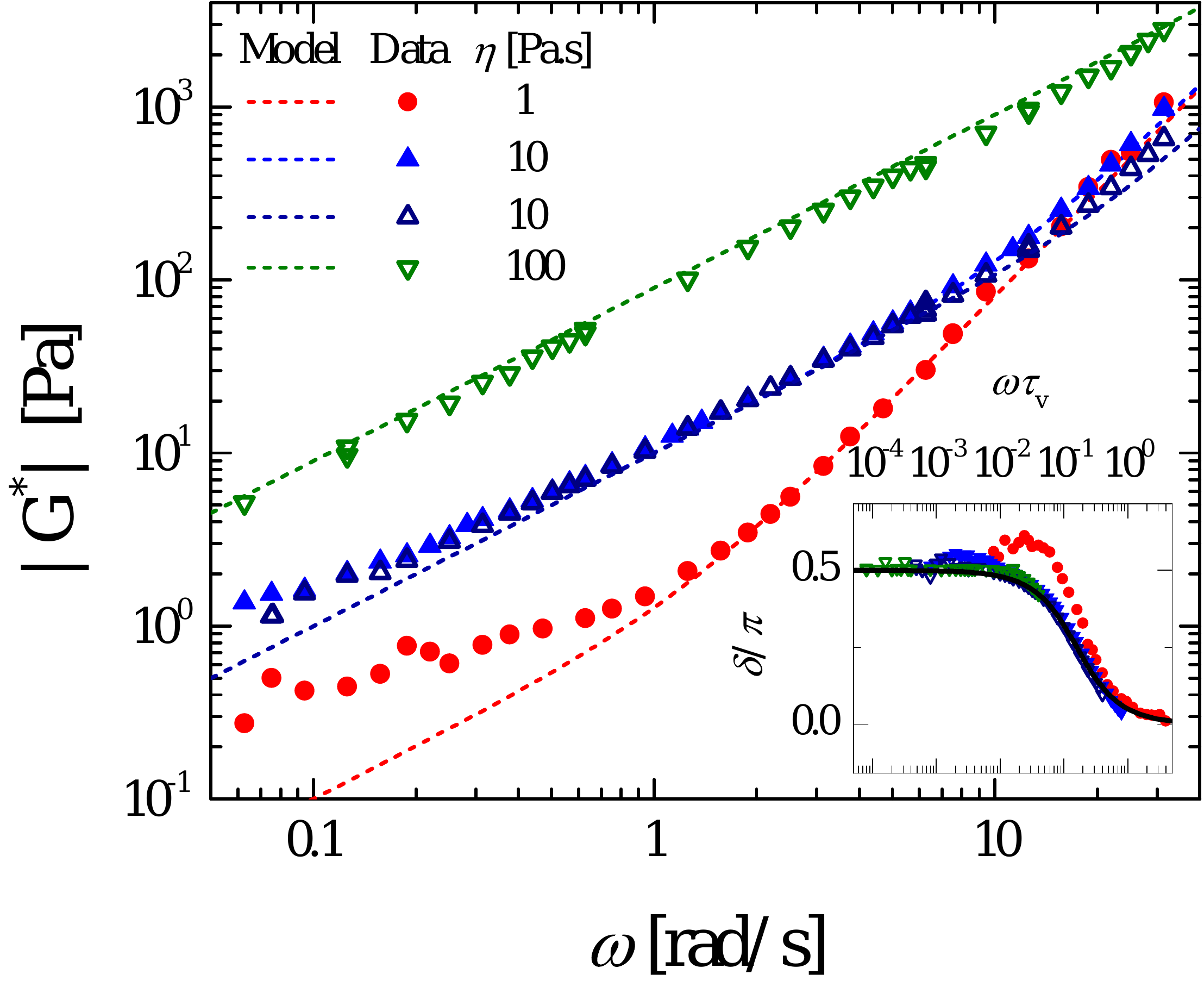}
  \caption{Main graph: magnitude of the complex modulus of Newtonian viscous fluids measured by oscillatory rheology, as a function of angular frequency. Solid (open) symbols are experiments performed with a gap $e=300~\mu \mathrm{m}$ ($e=600~\mu \mathrm{m}$). The data are labelled by the nominal viscosity. The dashed lines are the theoretical predictions, Eq.~\ref{eqn:viscous_moduli}, using the nominal viscosity and the inertia term issued from measurements on the empty cell. Inset: phase $\delta$, as a function of angular frequency normalized by the visco-inertial time $\tau_v$. The line is the theoretical prediction, Eq.~\ref{eqn:viscous_moduli}.}
  \label{fig:viscous_oscill}
\end{figure}

Measurements for oscillating stresses have been performed with the same 3 samples and two different values of the gap: $e=300~\mu$m and $e=600~\mu$m (solid and open symbols of fig.~\ref{fig:viscous_oscill}, respectively). The stress amplitude was chosen in order to keep the strain amplitude $|\tilde{\gamma}_\omega|$ fixed at $20\%$ for all $\omega$. In fig.~\ref{fig:viscous_oscill} the magnitude of the complex modulus $\left|G^*\right|$ (main plot) and its phase $\delta$ (inset) are plotted as a function of angular frequency, together with the theoretical expectations given by Eq.~\ref{eqn:viscous_moduli}. The data are in very good agreement with the theory for applied stresses with $|\tilde{\sigma}_\omega| > 1~\mathrm{Pa}$. By contrast, when lower stresses are applied to obtain the same strain amplitude ($20\%$ $\gamma$) --i.e. for the less viscous samples and the lowest frequencies--, deviations are observed due to the residual friction discussed above. Some deviations are also observed at the highest $\omega$, due to the response of the current generator and voice coil, as mentioned in Sec.~\ref{sec:empty}. For the more viscous sample ($\eta = 100~\mathrm{Pa~s}$), no inertial regime is observed: $|G^*|$ scales as $\omega$, as expected for a Newtonian fluid. This is consistent with the fact that for this sample inertial effects should become significant only for $\omega \gtrsim 1/\tau_v  = 100~\mathrm{rad~s}^{-1}$, which is beyond the range of probed frequencies. For the sample with intermediate viscosity ($\eta = 10~\mathrm{Pa~s}$), a crossover between $|G^*| \sim \omega$ (viscous regime) and $|G^*| \sim \omega^2$ (inertia-dominated regime) is observed at $\omega \approx 10-15~\mathrm{rad~s}^{-1}$, consistent with $1/\tau_v  = 10-20\mathrm{~s}^{-1}$ (depending on $e$). Finally, for the less viscous sample we estimate $1/\tau_v  = 2\mathrm{~s}^{-1}$: for this sample, all data for which friction is negligible lay in the high frequency regime $\omega \tau_v > 1$ where inertia dominates, leading to the observed $|G^*| \sim \omega^2$ scaling. The inset of fig.~\ref{fig:viscous_oscill} shows the argument of the complex modulus as a function of the normalized frequency $\omega \tau_v$. The data nicely exhibit the transition between the low and high frequency regimes predicted by Eq.~\ref{eqn:viscous_moduli}, characterized by $\delta = \pi/2$ (viscous regime) and $\delta = 0$ (inertial regime), respectively. The discrepancy seen for the fluid with $\eta = 1~\mathrm{Pa~s}$ is again due to residual friction.

\subsection{Purely elastic solid}

\subsubsection{Theoretical background}
\label{sec:elastic_theo}
The constitutive law for a purely elastic sample is $\sigma = G \gamma$, with $G$ the elastic shear modulus. Equation~\ref{eqn:force_balance_unitarea} then becomes
\begin{equation}
	I \ddot{\gamma} + G \gamma = \sigma^{(ext)}\,,
	\label{eqn:elastic_equation}
\end{equation}
whose solution involves the characteristic frequency $\Omega=\sqrt{G/I}$, related to an inertial time scale. Inertia is expected to dominate on time scales smaller than $\Omega^{-1}$, whose typical value for our setup, using $M=0.3$ kg, $e=300~\mu$m, and $A=2~\mathrm{cm}^2$, is given by $\Omega^{-1}~\mathrm{[s]} \approx 0.7/ \sqrt{G~\mathrm{[Pa]}}$.

In a step stress experiment with $\sigma^{(ext)} = \sigma_0\Theta(t)$, the solution to Eq.~\ref{eqn:elastic_equation} is
\begin{equation}
	\gamma(t)=\gamma_\infty \left[1 - \cos\left(\Omega t\right)\right]\,,
	\label{eqn:elastic_general_sol_step}
\end{equation}
with $\gamma_\infty = \sigma_0/G$. In the inertial regime $\Omega t \ll 1$ one finds the characteristic quadratic time dependence of the strain, $\gamma(t)\approx  \frac{1}{2}\frac{\sigma_0}{I}t^2$, while at later times the $\cos\left(\Omega t\right)$ term leads to a distinctive ``ringing'' behavior. Note however that when friction is included, the strain oscillations eventually are damped.

For an oscillatory imposed stress, the solution to Eq.~\ref{eqn:elastic_equation} reads
\begin{equation}
	\gamma(t) = \tilde{\gamma}_\omega\left[e^{i \omega t} - \cos\left(\Omega t\right) - i\frac{\omega}{\Omega} \sin \left(\Omega t\right) \right]\,,
	\label{eqn:elastic_general_sol_oscill}
\end{equation}
with $\tilde{\gamma}_\omega = \frac{\tilde{\sigma}_\omega}{G}\left[1-\left(\frac{\omega}{\Omega}\right)^2\right]^{-1}$. This corresponds to a complex modulus whose magnitude and phase are
\begin{equation}
\label{eqn:solid_moduli}
\begin{array}{r@{}l}
		|G^*| = G \left[1-\left(\frac{\omega}{\Omega}\right)^2\right]  \\
		\delta =  0\,.
\end{array}
\end{equation}
At small times $t \ll \Omega^{-1}$ we have again $\gamma(t) \approx \tilde{\gamma}_\omega\left( e^{i\omega t}-1-i \omega t\right)$, whereas at large $t$ the strain exhibits oscillations at the frequency $\Omega$ associated to inertia, superimposed to the elastic response at the forcing frequency $\omega$: their amplitude is $|\tilde{\sigma}_\omega|/G$ and they are in phase with the driving force, a distinctive feature of elasticity.

\subsubsection{Experimental tests}

\begin{figure}[h]
\centering
  \includegraphics[width=\columnwidth,clip]{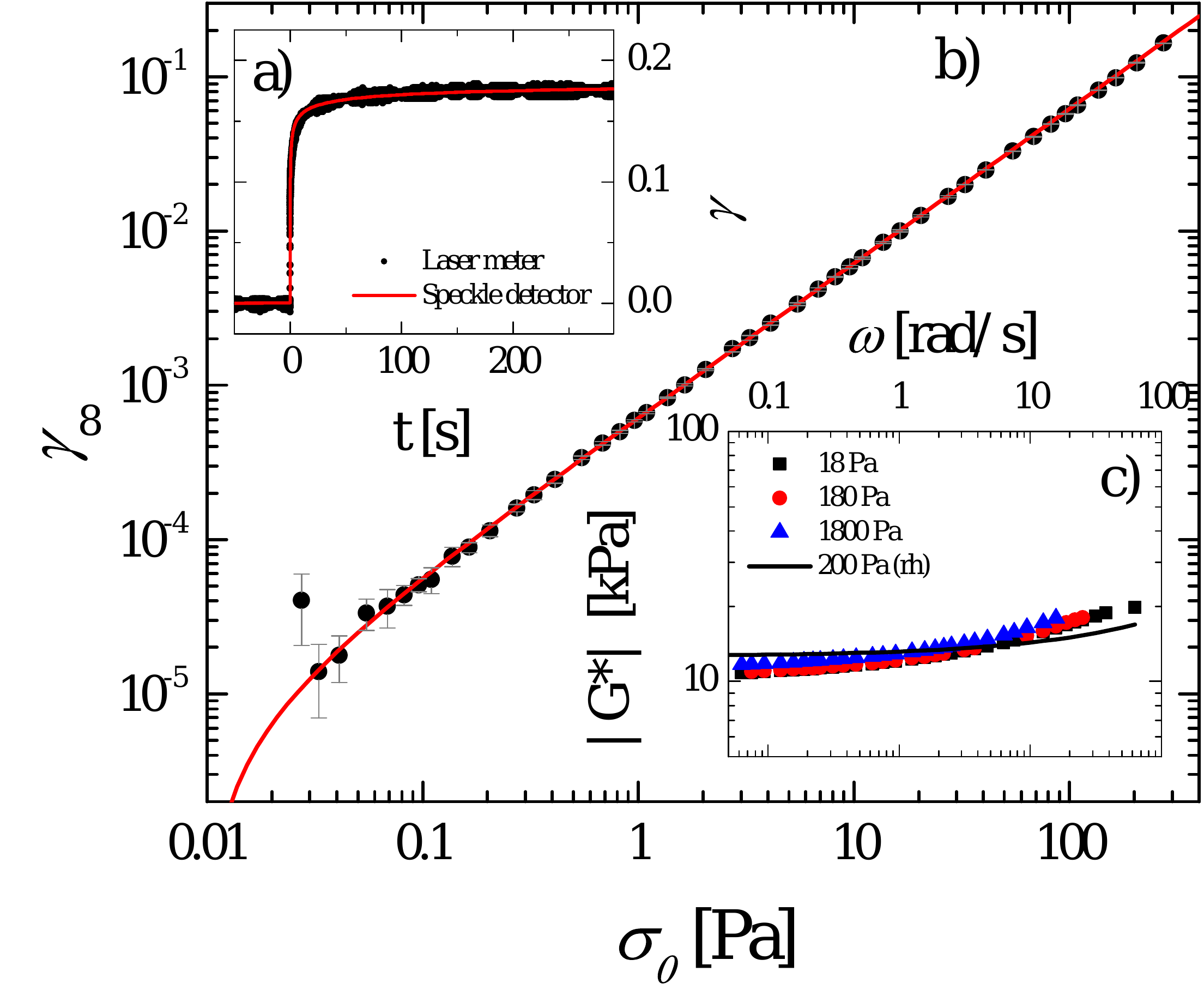}
  \caption{(a) Time evolution of the strain following a step stress applied at $t=0$, measured with both the commercial laser meter (black dots) and the home made optical setup (red line). (b)  Response to a step stress. The asymptotic strain $\gamma_\infty$ is plotted against the amplitude of the applied step stress. Data are averaged over $8$ experiments, the error bars are the standard deviation over the experiment repetitions. The solid line is an affine fit, yielding an elastic modulus $G = 1620$ Pa. (c) Frequency dependence of the elastic modulus in an oscillatory test, as measured in the custom shear cell at different stress amplitudes (symbols) and in a commercial rheometer (line).}
  \label{fig:elastic}
\end{figure}

We use polydimethylsiloxane (PDMS; Sylgard 184 by Dow Corning), a popular elastomer whose elastic modulus can be conveniently tuned by varying the crosslink density~\cite{gutierrez_measurements_2011-1}, as a model elastic solid. Using a commercial rheometer, we check that the linear regime extends up to at least $\gamma = 40\%$, larger than the largest strain probed in our tests. A sample with a crosslinker/base volume ratio of 1:60 is used for the  step stress measurements. In order to maximize the adhesion to the glass plates, the sample is prepared in situ: after adding the crosslinker, the fluid solution is  placed between the two plates and cured in an oven at $90^\circ$C for 90 minutes. The two plates with the cured sample are then mounted on the shear cell. During this operation, care is taken not to stress nor damage the sample. Control measurements are run in a commercial rheometer, with the sample cured in situ under similar conditions. We perform a series of step stress experiments, with a gap $e = 570~\mu$m (as measured before the experiment) and a sample surface $A=3.3$ cm$^2$. Stresses of various amplitude $\sigma_0$ are applied, the resulting strain being recorded over time.  Figure~\ref{fig:elastic}a shows a typical time series of strain data acquired simultaneously with the commercial laser sensor (symbols) and with our custom made optical setup (line), for $\sigma_0 = 275~\mathrm{Pa}$: note the excellent agreement between the two devices. After an initial transient, lasting a few seconds, $\gamma$ reaches a constant value $\gamma_\infty$ dictated by the elastic modulus. Note that the oscillations due to inertia predicted by Eq.~\ref{eqn:elastic_general_sol_step} are not seen here, indicating that the (small) dissipation of PDMS is sufficient to overdamp them. In fig.~\ref{fig:elastic}b, $\gamma_\infty$ estimated by averaging $\gamma(t)$ over $5$ minutes is plotted as a function of the imposed stress. Each data point is an average over $8$ experiments, the error bars representing data dispersion. The experimental data are very well fitted by an affine law (red line): $\gamma_\infty=(\sigma_0-\sigma_{off})/G$, with $G=1620$ Pa the elastic modulus of the sample and $\sigma_{off}=13$ mPa an offset stress due to friction. We note that $\sigma_{off}$ is smaller than the typical friction stress estimated from the experiments with an empty cell, which is of the order of $1~\mathrm{Pa}$. This discrepancy suggests that the exact value of the friction term depends sensitively on several factors (cleanness of the air bearing, alignment with respect to the vertical direction...) and that $1~\mathrm{Pa}$ should be regarded as a higher bound on its magnitude.

A second series of experiments is performed by imposing oscillating stresses. The sample is prepared at a higher crosslinker density (1:40 v/v), with $e=590~\mu$m and $A=2.5$ cm$^2$. After curing the PDMS as described above, an oscillating stress is applied and the amplitude of the strain oscillations is extracted from a sinusoidal fit to $\gamma(t)$. Figure~\ref{fig:elastic}c shows the frequency dependent elastic modulus obtained from $G(\omega)=|\sigma_\omega/\gamma_\omega|$, as a function of angular frequency. Experiments performed at different stress amplitudes spanning two decades within the linear regime nicely overlap in the whole frequency range. Additionally, data taken in the custom shear cell overlap reasonably well with data acquired in a commercial rheometer for a similar sample, thus validating the new setup. The slight differences that are seen are most likely due to the different protocols for preparing and loading the sample in the two devices.

\subsection{Maxwell fluid}

\subsubsection{Theoretical background}

As an example of a viscoelastic fluid, we consider a Maxwell fluid, for which $\dot{\gamma} = \dot{\sigma}/G + \sigma/\eta$. Here $G$ is the plateau modulus and $\eta = G \tau_M$ the viscosity, with $\tau_M$ the Maxwell relaxation time. Using this constitutive law, the equation of motion, Eq.~\ref{eqn:force_balance_unitarea}, yields:
\begin{equation}
	I\frac{d^2\dot{\gamma}}{dt^2} + \frac{I}{\tau_M}\frac{d\dot{\gamma}}{dt} + G \dot{\gamma} = \dot{\sigma}^{(ext)} + \frac{\sigma^{(ext)}}{\tau_M}\,,
	\label{eqn:maxwell_speed_diff_eq}
\end{equation}
whose solutions again involve the characteristic frequency $\Omega = \sqrt{G/I}$.

The general solution for a step stress is
\begin{equation}
	\gamma(t) = \frac{\sigma_0}{G} \left(c_0 + \frac{t}{\tau_M} + c_+ e^{-\lambda_+ t} + c_- e^{- \lambda_- t}\right)\,,
\end{equation}
where $c_0 = 1- \frac{1}{\tau_M^2\Omega^2}$, $c_\pm = -\frac{1}{2}\left(c_0 \mp \frac{c_0+2}{\sqrt{1-4\tau_M^2\Omega^2}}\right)$, and $\lambda_\pm=\frac{1}{2\tau_M} \left(1 \pm \sqrt{1-4\Omega^2\tau_M^2}\right)$.
The regime of slowly relaxing Maxwell fluids (as compared to the inertial time) corresponds to $\Omega\tau_M \ll 1$. In this limit, one recovers $\gamma \approx \sigma t^2/(2I)$, as for a purely viscous fluid (see Eq.~\ref{eqn:viscous_motion}). In the opposite limit $\Omega\tau_M \gg 1$, the solution is:
\begin{equation}
	\gamma(t) = \frac{\sigma_0}{G} \left[1 + \frac{t}{\tau_M} - \exp \left (- \frac{t}{2\tau_M}\right)\cos(\Omega t)\right]\,,
	\label{eqn:maxwell_creep_approx}
\end{equation}
for which three regimes may be distinguished. For $t \ll \Omega^{-1}$, inertia dominates and $\gamma = \sigma_0 t^2 /(2I)$. For, $\Omega^{-1} \ll t \ll \tau_M$, the typical oscillations due to the elastic part of the sample response are observed: $\gamma(t) \approx \frac{\sigma}{G} \left[1- \cos(\Omega t)\right]$. Finally, at long times $t \gg \tau_M$ the sample flows as a purely viscous fluid: $\gamma(t) \approx \sigma_0 t /\eta$.

For the sake of completeness, we report here the equations for an applied oscillating stress, although in the following we test the Maxwell fluid only in step stress experiments. Focussing on the complex modulus calculated from the oscillating part of the cell response, one finds:
\begin{equation}
\label{eqn:maxwellfriction}
\begin{array}{r@{}l}
|G^*| = G \frac{\left(1-\frac{\omega^2}{\Omega^2}\right)^2+\frac{\omega^2}{\Omega^4\tau_M^2}}{\sqrt{\left(1-\frac{\omega^2}{\Omega^2}-\frac{1}{\tau_M^2\Omega^2}\right)^2+\frac{1}{\omega^2\tau_M^2}}} \\
\tan\delta = \frac{1}{\omega\tau_M}\frac{\Omega^2}{\Omega^2-\omega^2-\frac{1}{\tau_M^2}}\,.
\end{array}
\end{equation}
In the $\Omega \gg \tau_M^{-1}$ and $\Omega \gg \omega$ limit where inertia is negligible Eq.~\ref{eqn:maxwellfriction} simplifies to yield the usual expressions for a Maxwell fluid:
\begin{equation}
\begin{array}{r@{}l}
		|G^*| = G \left( 1+\frac{1}{\omega^2\tau_M^2} \right) ^{-\frac{1}{2}} \\
		\tan\delta = \frac{1}{\omega\tau_M} \,.
\end{array}
\end{equation}

\subsubsection{Experimental tests}

\begin{figure}[h]
\centering
  \includegraphics[width=\columnwidth,clip]{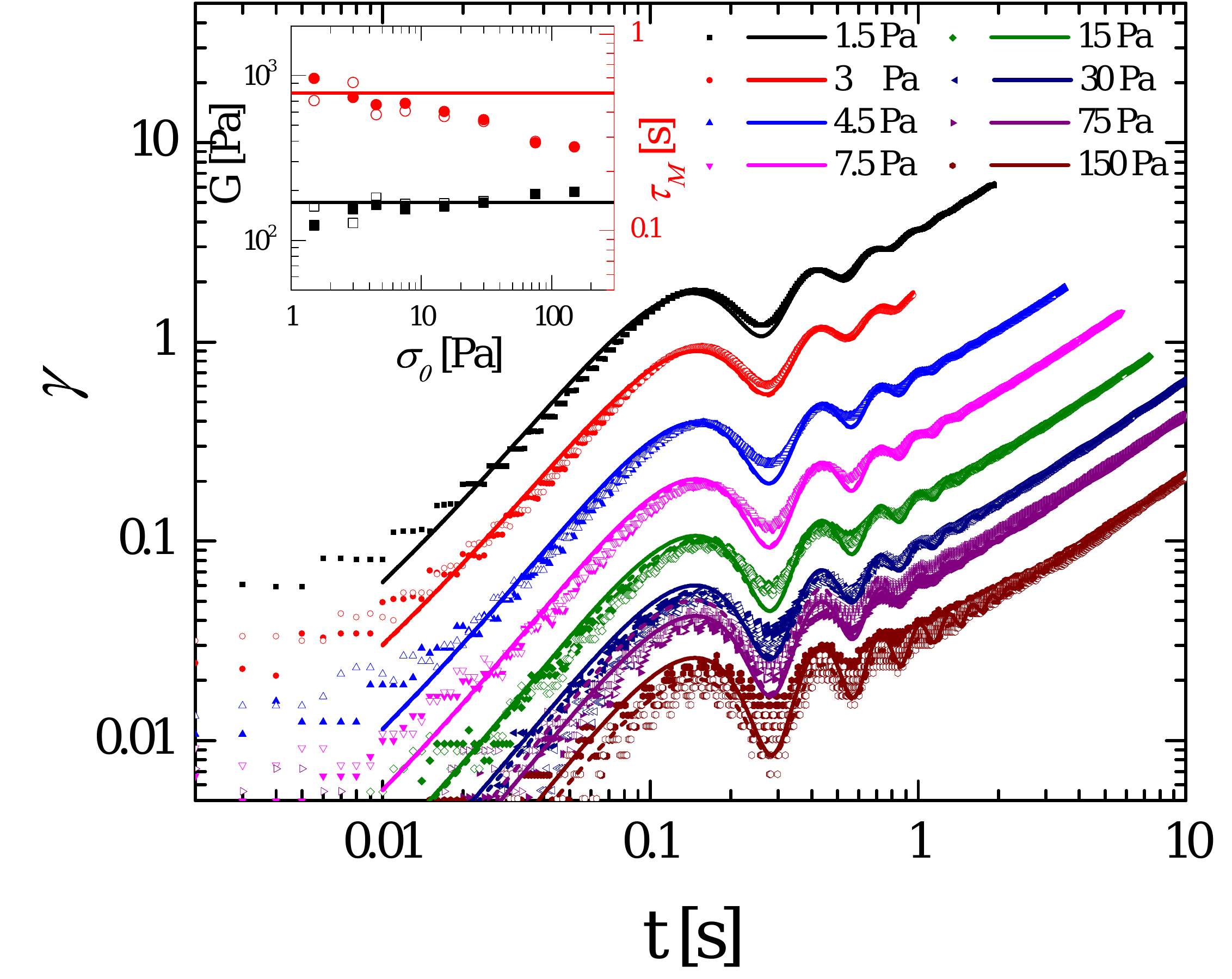}
  \caption{Main plot: strain for a Maxwell sample as a function of time following a step stress with amplitude $\sigma_0$ as given by the labels. Open and filled symbols represent two consecutive measurements on the same sample; lines are best fits to the data using Eq.~\ref{eqn:maxwell_creep_approx}, where the fitting parameters are the characteristic relaxation time, $\tau_M$, and the plateau modulus, $G$.  Inset: Maxwell parameters extracted from the fits of the experimental creep curves shown in the main plot, plotted against the applied stress. Squares and left axis: plateau modulus; circles and right axis: relaxation time. The symbols are experimental data obtained in the custom shear cell, the lines are $\tau_M$ and $G$ as measured in a commercial rheometer, in a frequency sweep test at $10\%$ strain amplitude.}
  \label{fig:maxwell_step}
\end{figure}

As a model Maxwell fluid, we use a self-assembled transient network, comprising surfactant-stabilized microemulsions reversibly linked by triblock copolymers, described in detail in~\cite{michel_percolation_2000}. After loading the sample in the shear cell, we wait 5 minutes before applying a step stress, in order to let the sample fully relax. Similarly, after each creep experiment a waiting time of $2$ minutes is applied before starting the next one. 
Creep experiments are performed for different stress amplitudes, ranging from $1.5$ to $150$ Pa. The time evolution of the strain amplitude is shown in fig.~\ref{fig:maxwell_step} for several applied stresses $\sigma_0$. The data are very well fitted by the theoretical expression (Eq.~\ref{eqn:maxwell_creep_approx}) using the viscoelastic parameters of the Mawxell fluid, $\tau_M$ and $G$, as fitting parameters. As shown in the inset of fig.~\ref{fig:maxwell_step}, we find over the whole range of applied stresses a nearly constant value of the elastic plateau ($G=165 \pm 20$ Pa), in excellent agreement with the one measured in a frequency sweep test in a conventional rheometer (180 Pa). A somehow poorer agreement between the shear cell data and conventional rheometry is seen for the characteristic relaxation time, which is found to decrease by a factor of almost $2$ as the applied stress increases. This is most likely due to the fact that, for this sample, the relaxation time $\tau_M$ is close to the characteristic ringing time due to inertia, $2\pi/\Omega \approx 0.3~\mathrm{s}$, which makes it difficult to independently and reliably retrieve the two times when fitting the data to Eq.~\ref{eqn:maxwell_creep_approx}.


\section{Dynamic Light Scattering under shear}
\label{sec:dls}

We couple the shear cell to a custom made small angle light scattering setup~\cite{tamborini_multiangle_2012} and measure the microscopic dynamics of a colloidal suspension of Brownian particles, both at rest and under shear. The setup uses a CMOS camera as a detector; the microscopic dynamics is quantified by the intensity correlation function $g_2-1$:~\cite{berne_dynamic_1976,duri_time-resolved-correlation_2005}
\begin{equation}
\label{eqn:dls}
g_2(\mathbf{q},\tau)-1 = \left < \frac{\left<I_p(t+\tau)I_p(t)\right>_{\mathbf{q}}}{\left<I_p(t)\right>_{\mathbf{q}}\left<I_p(t+\tau)\right>_{\mathbf{q}}}\right>_t-1\,.
\end{equation}
Here, $I_p$ is the time-dependent intensity measured by the $p-th$ pixel, $<\cdot \cdot \cdot >_t$ denotes an average over the experiment duration and $<\cdot \cdot \cdot >_{\mathbf{q}}$ is an average over a set of pixels corresponding to a small solid angle centered around the direction associated to a scattering vector $\mathbf{q}$, whose magnitude is $q = 4\pi n \lambda^{-1} \sin(\theta/2)$, with $n$ the solvent refractive index, $\lambda = 532$ nm the laser wavelength and $\theta$ the scattering angle.

The intensity correlation function is directly related to the microscopic dynamics: for particles undergoing Brownian motion, $g_2(\mathbf{q}, \tau) - 1 = \exp(-2Dq^2\tau)$, where $D$ is the particle diffusion coefficient~\cite{berne_dynamic_1976}. Under shear, the relative motion due to the affine displacement also contributes to the decay of $g_2-1$. Neglecting hydrodynamic interactions and assuming an homogeneous shear flow in the $xz$ plane ($\hat{u}_x$ and $\hat{u}_z$ being the velocity and the velocity gradient direction, respectively), the particle dynamics are described by the displacement field  $\Delta\mathbf{r}(\mathbf{r}, \dot{\gamma}, \tau)=\Delta\mathbf{r}^{\mathrm{(diff)}}(\tau) + \Delta\mathbf{r}^{\mathrm{(aff)}}(\mathbf{r}, \dot{\gamma},\tau)$, where $\Delta\mathbf{r}^{\mathrm{(aff)}}(\mathbf{r}, \dot{\gamma},\tau) = \dot{\gamma}\tau(\mathbf{r} \cdot \hat{u}_z)\hat{u}_x$ is the affine shear field, whereas $\Delta\mathbf{r}^{\mathrm{(diff)}}(\tau)$ describes Brownian diffusion. Assuming that the two contributions are uncorrelated, the intensity correlation function can be expressed as~\cite{Stefano_in_preparation}:
\begin{equation}
\label{eqn:brownshear}
	g_2(\mathbf{q}, \dot{\gamma}, \tau) - 1 = \sinc^2\left(\frac{e}{2} q_x\dot{\gamma}\tau\right) \exp\left(-2Dq^2\tau\right)\,,
\end{equation}
where $q_x=\mathbf{q}\cdot\hat{u}_x$ is the projection of the scattering vector along the velocity direction. Note that no contribution of the applied shear on the dynamics is expected when selecting a scattering vector $\mathbf{q}=q_y\hat{u}_y$ parallel to the vorticity direction.

\begin{figure}[h]
\centering
  \includegraphics[width=\columnwidth,clip]{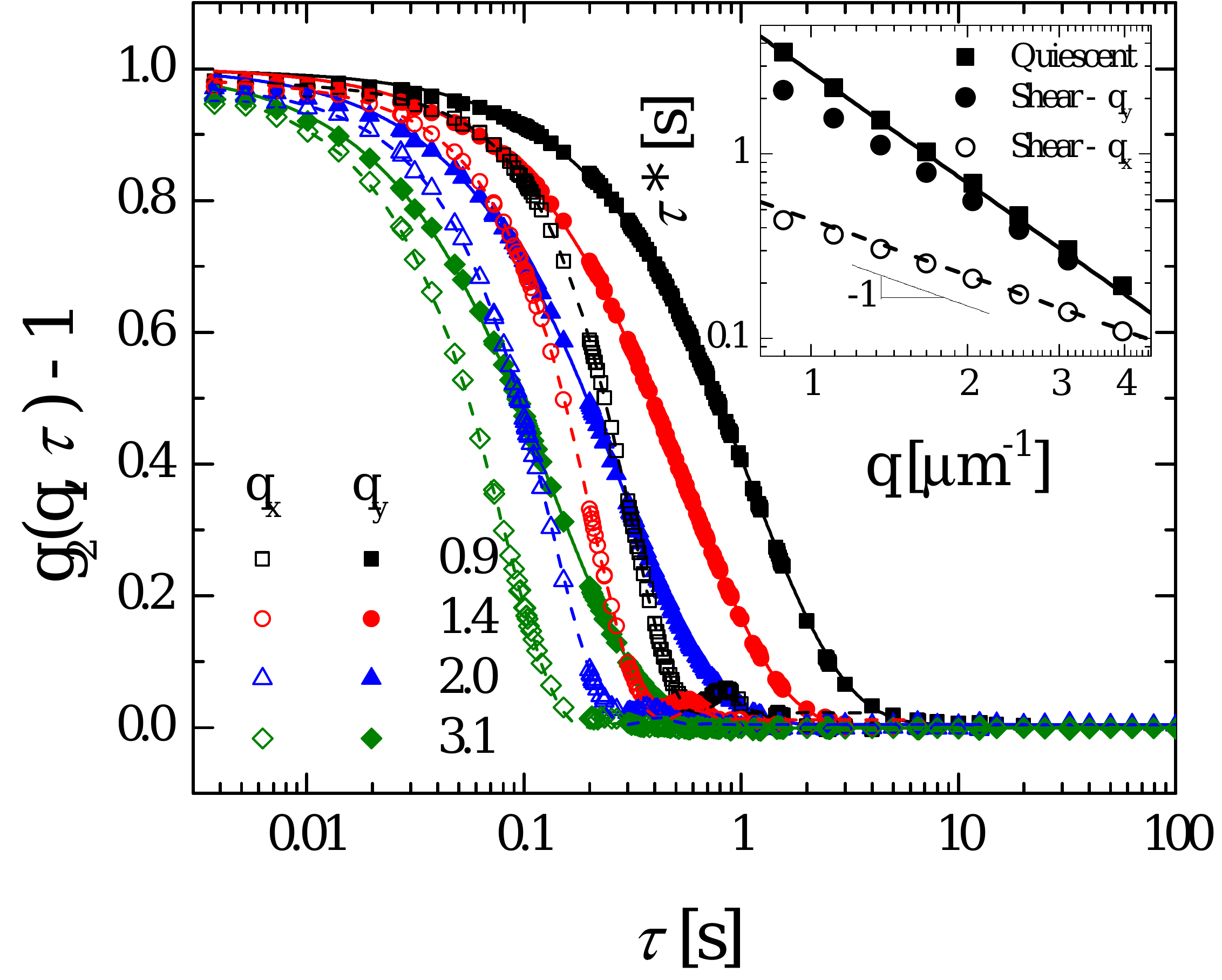}
  \caption{Main plot: representative intensity correlation functions for a sheared suspension of Brownian particles, for scattering vectors parallel ($q_x$, open symbols) and perpendicular ($q_y$, solid symbols) to the shear direction. Data are fitted with Eq.~\ref{eqn:brownshear} (dashed and solid lines for $q_x$ and $q_y$, respectively) with $D$ and $\dot{\gamma}$ as fitting parameters. Inset: characteristic time $\tau^*$ as defined in the text for the two orientations of $\mathbf{q}$ and for a quiescent sample, as a function of the scattering vector. The lines are the predictions of Eq.~\ref{eqn:brownshear}, using the nominal values of the diffusion coefficient, $D=0.36~\mu\mathrm{m}^2\mathrm{s}^{-1}$, and the shear rate, $\dot{\gamma}=0.03~\mathrm{s}^{-1}$.}
  \label{fig:brownshear}
\end{figure}

We measure the dynamics of a suspension of polystyrene particles with diameter $2a=1.2\mu\mathrm{ m}$ (Invitrogen Molecular Probes), suspended at a weight fraction of 0.01\% in a 1:1 v/v mixture of $\mathrm{H_2O}$ and $\mathrm{D_2O}$ that matches the density of polystyrene, to avoid sedimentation. A time series of images of the scattered light is acquired using the scheme described in~\cite{philippe_efficient_2016}, both at rest and while imposing a constant shear rate $\dot{\gamma}=0.03s^{-1}$. The intensity correlation function is calculated using Eq.~\ref{eqn:dls}, for several $\mathbf{q}$ vectors both along the velocity and the vorticity direction ($q_x$ and $q_y$, respectively). Figure \ref{fig:brownshear} shows representative correlation functions; for a given magnitude of the scattering vector, the decay time is shorter when $\mathbf{q}$ is oriented parallel to the shear velocity (open symbols), reflecting faster dynamics due to the contribution of the affine shear field. The lines are fits to the data using Eq.~\ref{eqn:brownshear}: for both $\mathbf{q}$ orientations, an excellent agreement is seen between the data and the model. We extract from the fits the characteristic relaxation time $\tau^*$, defined by $g_2(\mathbf{q}, \tau^*) - 1 = \exp(-2)$, and plot $\tau^*$ \textit{vs} $q$ in the inset of Fig.~\ref{fig:brownshear}. For $\mathbf{q}=q_y\hat{u}_x$, $\tau^*$ is in excellent agreement with the theoretical prediction, dotted line; its $q$ dependence is very close to a $q^{-1}$ scaling, indicative of ballistic motion and suggesting that over the range of probed $q$ vectors and time scales the displacement due to the affine deformation is significantly larger than that due to Brownian motion. This is confirmed by inspecting $\tau^*(q)$ for a quiescent suspension (solid squares and solid line): for all $q$, the relaxation time due to Brownian motion is larger than that under shear. The solid circles are the relaxation time of the dynamics in the vorticity direction while shearing the sample. Overall, $\tau^*(q_y)$ closely follows the data for the quiescent suspension, demonstrating that the affine and non-affine components of the particle dynamics can be resolved by dynamic light scattering. Only at the smallest $q$ vector do the data exhibit some deviations with respect to purely Brownian dynamics: most likely, this discrepancy stems from slight deviations of the velocity field from the ideal one, due to the finite size of the sample.

\section{Conclusion}
\label{sec:conclusions}
We have presented a custom-made stress-controlled shear cell in the linear translation plate-plate geometry, which can be coupled to both a microscope and a static and dynamic light scattering apparatus. The setup has been successfully tested on a variety of samples representative of simple fluids, ideal solids and viscoelastic fluids; its coupling to dynamic light scattering measurements has been illustrated by measuring the dynamics of Brownian particles in a suspension at rest or under shear. The main features of the cell include a gap that can be adjusted down to $100\um$ keeping the parallelism to within 0.3 mrad, the acquisition of strain data at up to 1000 Hz with very good precision (typically better than 0.01\%) and for strains as large as $10^4\%$, and the possibility of imposing a user-defined, time-varying stress, ranging from 0.1 Pa (limited by residual friction) to 10 kPa, with a frequency bandpass $0 \le \omega \le 4~\mathrm{rad s^{-1}}$. Most of these specifications meet or even exceed those of a commercial rheometer. Among the limitations intrinsic to the setup, the most stringent is probably inertia, which restricts viscosity measurements to $\eta \gtrsim 1~\mathrm{Pa~s}$. Other limitations in the current implementation could be improved: for example, the upper stress limit could be pushed to 100 kPa by changing the magnetic actuator and the cutoff frequency $\omega \approx 4~\mathrm{rad s^{-1}}$ could be increased by modifying the current generator design. Finally, we remind that care must be taken in aligning both the plate parallelism and the horizontality of the setup, which makes its use less straightforward than that of a commercial rheometer. These drawbacks are more than offset by the advantages afforded by our shear cell: the possibility of easily changing the plates, e.g. to optimize them against slippage or to replace them when the surface quality doesn't meet anymore the stringent requirements for microscopy or light scattering; the open design that leaves full access on both sides, thus making it suitable for both inverted and upright microscopes and for small angle light scattering; the linear translation geometry that insures uniform stress in the (optically optimal) parallel plate configuration; the flexibility provided by the choice of the orientation in the vertical or horizontal plane; and, last but not least, the reduced cost as compared to a commercial rheometer (about 4000 Euros including a PC and an air filter for the air bearing rail).

\begin{acknowledgments}
We thank J. Barbat and E. Alibert for help in instrumentation and S. Arora for providing us with the bridged microemulsion sample. Funding from the French ANR (project FAPRES, ANR-14-CE32-0005-01) and the  E.U. (Marie Sklodowska-Curie ITN ‘Supolen’, no 607937) is gratefully acknowledged.
\end{acknowledgments}

\textbf{Bibliography}\\
\nocite{*}

%

\end{document}